\documentclass[apj]{emulateapj}
\pdfoutput=1
\usepackage{color}
\usepackage{times}
\usepackage{graphicx}
\usepackage{amssymb,amsmath}

\providecommand{\abs}[1]{\big \vert#1 \big\rvert}

\slugcomment{Accepted to ApJ, July 29 2013}

\shorttitle{First Order Resonance Overlap in Two Planet Systems}
\shortauthors{Deck et al.}

\begin{document}

\title{First order resonance overlap and the stability of close two planet systems}

\author{Katherine M. Deck\altaffilmark{1,3}, Matthew Payne\altaffilmark{2}, Matthew J. Holman\altaffilmark{2}}

\altaffiltext{1}{Department of Physics and Kavli Institute for Astrophysics and Space Research,
Massachusetts Institute of Technology, 77 Massachusetts Ave., Cambridge, MA 02139}
\altaffiltext{2}{Harvard-Smithsonian Center for Astrophysics, 60 Garden St., Cambridge, MA 02138}
\altaffiltext{3}{Corresponding author: kdeck@mit.edu}

\begin{abstract}
Motivated by the population of observed multi-planet systems with orbital period ratios $1< P_2/P_1 \lesssim 2$, we study the long-term stability of packed two planet systems. The Hamiltonian for two massive planets on nearly circular and nearly coplanar orbits near a first order mean motion resonance can be reduced to a one degree of freedom problem \citep{Sessin,WisdomCtB,Lemaitre}. Using this analytically tractable Hamiltonian, we apply the resonance overlap criterion to predict the onset of large scale chaotic motion in close two planet systems. The reduced Hamiltonian has only a weak dependence on the planetary mass ratio $m_1/m_2$, and hence the overlap criterion is independent of the planetary mass ratio at lowest order. Numerical integrations confirm that the planetary mass ratio has little effect on the structure of the chaotic phase space for close orbits in the low eccentricity ($e\lesssim 0.1$) regime. We show numerically that  orbits in the chaotic web produced primarily by first order resonance overlap eventually experience large scale erratic variation in semimajor axes and are therefore Lagrange unstable.  This is also true of the orbits in this overlap region which satisfy the Hill criterion. As a result, we can use the first order resonance overlap criterion as an effective stability criterion for pairs of observed planets.  We show that for low mass ($ \lesssim 10 M_{\oplus}$) planetary systems with initially circular orbits the period ratio at which complete overlap occurs and widespread chaos results lies in a region of parameter space which is Hill stable. Our work indicates that a resonance overlap criterion which would apply for initially eccentric orbits likely needs to take into account second order resonances.  Finally, we address the connection found in previous work between the Hill stability criterion and numerically determined Lagrange instability boundaries in the context of resonance overlap.

\end{abstract}
\section{INTRODUCTION}
 Observational results from the \textit{Kepler} mission have revealed a large population of multi-planet systems with nearly circular and nearly coplanar orbits \citep{Batalha,Fabrycky,FM1}. The distribution of the orbital separation between adjacent pairs of planets, expressed either in terms of the observed period ratios or in terms of the inferred ``dynamical spacing" (mutual Hill radii), encodes a significant amount of information regarding how these systems formed and dynamically evolved \citep{Fabrycky,FM2}. One interesting question regarding the orbital separations of planets is whether or not the inferred dynamical spacing distribution is dictated by orbital stability requirements or by planetary formation processes. In order to address this, it is important to understand in a theoretical manner the long-term stability of close two-planet systems.

Addressing analytically the question of how small an orbital separation two planets can have while remaining  ``long-lived" is challenging, as even systems of two planets can be chaotic, and chaotic orbits cannot be described in terms of simple  functions. In general, to make analytic progress, researchers must turn to more global methods, rather than the study of individual initial conditions.

It has been proven that the conservation of angular momentum and energy can constrain the motion in two planet systems in a way such that crossing orbits never develop \citep{Marchal261982,Nobili}. Orbits which do not cross will never lead to collisions between planets or strong gravitational scattering. As a result, one can use the integrals of the motion to prove that a two planet system is collisionally (Hill) stable. However, planetary orbits which fail this ``Hill criterion" and evolve in the region of phase space where crossing orbits are possible are not guaranteed to be unstable; satisfying the criterion is a sufficient but not necessary condition for collisional stability. 

Additionally, the Hill criterion gives no information about the long-term behavior of orbits that satisfy the criterion. Even though two bodies may never suffer strong encounters, repeated interactions can lead to a net transfer of angular momentum and energy between the two bodies that result in large, erratic, variations in the orbits. In particular, the semimajor axes can change considerably, and in some cases this leads to ejection of the outer body or a collision between the inner and central bodies. These orbits are referred to as Lagrange (or Laplace) unstable, even though they are stable in the Hill sense. Our definition of Lagrange instability is not restricted to cases in which ejection of the outer planet or collisions between the central body and the inner planet occur. 

Systems which are protected from this type of behavior have more constrained orbits that we call Lagrange long-lived. It has been found numerically that orbits which marginally satisfy the Hill criterion exhibit Lagrange instabilities, but that orbits which satisfy the Hill criterion by a larger amount are generally Lagrange long-lived \citep{BG1,BG2, MudrykandWu, Kepler36}. These studies found that the transition between Lagrange unstable orbits and Lagrange long-lived orbits, as the ``distance" from the Hill boundary grows, occurs very abruptly, over a narrow range in the orbital parameters. A theoretical explanation for this behavior is lacking, though the specific location of the transition point in terms of the planetary mass and the orbital parameters has been explored numerically \citep{BarnesandKopparapu,VerasMustill}. 

\citet{Gladman} numerically studied the effectiveness of the Hill criterion as a stability criterion and found that for low mass planets on initially circular orbits, the Hill criterion was effectively $\mathit{necessary}$ for stability, while for more eccentric cases there seemed to be many orbits which failed the criterion but appeared to be long-lived. He also found that there existed a chaotic region of phase space which extended past the Hill boundary. A limited number of numerical experiments with equal mass planets indicated that the width of this region scaled as $\epsilon_p^{2/7}$, where $\epsilon_p$ is the total mass of the planets, relative to the star. The specific power of $2/7$ is motivated theoretically by analytic work on first-order resonance overlap in the circular restricted three body problem \citep{Wisdom1980,DuncanQuinnTremaine}. However, it is not known how this result changes in the case of two massive planets on low eccentricity orbits, with arbitrary mass ratio, or if first order resonance overlap accounts for the chaotic zone discovered by Gladman.

 In this work, we study the stability of two-planet systems by extending the resonance overlap criterion to the full planetary problem. We seek to explain Gladman's numerical results and investigate the intriguing correlation between Hill stable and Lagrange long-lived orbits and the role resonance overlap plays in this relationship. 
 
 We also approach the problem ``globally" in that we do not evolve individual orbits for very long times. Instead, we focus on predicting where the chaotic regions of phase space lie. A single chaotic orbit will explore densely the entire extent of the chaotic zone available to it. As a result, if a chaotic zone encompasses regions of phase space where Lagrange-type instabilities can set in, any chaotic orbit within that region is Lagrange unstable. An example of this would be a chaotic zone which extended over a large range of period ratios. The problem of long-term stability in planetary systems can therefore be approached by identifying the regions of phase space where widespread chaos will be present, rather than actually following the long-term evolution of single trajectories.

In order to identify chaotic regions of phase space we make use of the resonance overlap criterion of \citet{Chirikov}; see also \citet{Walker}. This heuristic criterion provides an intuitive way to predict where in phase space global chaos appears\footnote{He or she who ``desires to reach a cherished islet of stability in the violent (and stochastic!) sea of nonlinear oscillations should not rely upon the beacons only", where ``the beacons" are the rigorous mathematical theorems of nonlinear dynamics \citep{Chirikov}. }, and has been used successfully to estimate regions of chaos in the restricted three body problem, in close three planet systems, in widely spaced two planet systems, and in explaining ejection of planets orbiting in binary star systems; see for example \citet{Lecar, Mardling, MudrykandWu,Quillen2011}.

We apply the resonance overlap criterion to a system with two massive planets on close orbits ($P_2/P_1 \lesssim 2$) which are nearly circular ( $ e \lesssim 0.1$) and nearly coplanar. In order to do so, we must first identify the dominant resonances in this regime. These are the first order mean motion resonances, which are important when the planetary period ratio $P_2/P_1 \sim (m+1)/m \leq 2$, where $m$ is a positive integer.  The distance in period ratio between the first order resonances shrinks as the planetary period ratio grows closer to unity, but the widths of the resonances (in terms of period ratio) do not shrink as quickly. Consequently, the first order resonances must overlap as the period ratio shrinks to unity, leading to chaotic motion. 

In Section \ref{sec:HamDevelop} we show how to reduce the first order resonance problem, at first order in the planetary eccentricities, to a one degree of freedom problem. We derive the resonance overlap criterion in Section \ref{sec:Resoverlap}. In Section \ref{sec:Compare}, we compare the predicted resonance widths and overlap criterion to results from numerical integrations. In Section \ref{sec:Discuss} we discuss the implications of resonance overlap for the long-term stability of close two planet systems.

\section{ANALYTIC REDUCTION OF THE HAMILTONIAN}\label{sec:HamDevelop}
Here we reduce the full Hamiltonian for two planets orbiting near a first order mean motion resonance with nearly circular and nearly coplanar orbits to a one-degree of freedom system with a single free parameter\citep{Sessin}. We follow a sequence of canonical transformations originally used by \cite{WisdomCtB} and \cite{Lemaitre} (and recently by \citet{Batygin}). Although this reduction of the Hamiltonian is not original work of ours, we include it here for consistency of notation throughout the text and to familiarize the reader.

The Hamiltonian governing the dynamics of a system of two planets of mass $m_i$ orbiting a much more massive star of mass $m_\star$, written in Jacobi coordinates $\bf{{r}}_i$ and momenta $\bf{{p}}_i$, takes the form
\begin{align}
{H} & = {H}_0 + {H}_1 \nonumber \\
{H}_0 & = {H}_{Kepler 1} + {H}_{Kepler 2} \nonumber \\
{H}_{Kepler 1} & =  \frac{{p}_1^2}{2\tilde{m}_1}   -\frac{G \tilde{M}_1 \tilde{m}_1}{\abs{\bf{{r}_1}}} \nonumber \\
{H}_{Kepler 2} & =  \frac{{p}_2^2}{2\tilde{m}_2}   -\frac{G \tilde{M}_2 \tilde{m}_2}{\abs{\bf{{r}_2}}} \nonumber \\
{H}_1 & = -G {m}_1 {m}_2\left(\frac{1}{\abs{ \bf{{r}_1} - \bf{{r}_2}}}  - \frac{ \bf{{r}_2} \cdot \mathbf{{r}_1}}{\abs{\bf{{r}_2}}^3} \right) +O(\epsilon^2),
\end{align}
where $\tilde{M}_2 = m_\star(m_\star+m_1+m_2)/(m_\star+m_1) $, $\tilde{M}_1 =  (m_\star+m_1) $, $\epsilon =$max($m_i/m_\star$), and $\tilde{m}_i =m_i+O(\epsilon)$ denote Jacobi masses. A subscript of $1$ refers to the inner planet, a subscript of $2$ refers to the outer planet. The interaction potential between the planets takes the form of the disturbing function \citep{MurrayDermott}. We set $\tilde{M}_i= m_\star$ and also ignore the difference between Jacobi and physical masses.  In the Keplerian piece ${H}_0$, this approximation corresponds to a change in the mean motions of the planets by order $\epsilon$, but this is negligible.

Instead of Cartesian coordinates and momenta, we use the Poincare canonical variables 
\begin{align}\label{var}
\Lambda_i &= m_i \sqrt{G m_\star a_i}  \nonumber \\
P_i &= m_i \sqrt{G m_\star a_i}  (1-\sqrt{1-e_i^2}) \nonumber \\
G_i & = m_i \sqrt{G m_\star a_i(1-e_i^2)}(1-\cos{I_i})
 \end{align} 
 and their conjugate angles $\lambda_i = \varpi_i+M_i$, $p_i = -\varpi_i $, and $g_i = -\Omega_i$. Here $a$, $e$, $I$, $\lambda$, $\Omega, \varpi$ and $M$ denote the (Jacobi) semimajor axis, eccentricity, inclination, mean longitude, longitude of ascending node, longitude of periastron  and mean anomaly of the osculating orbit.  When expressed in terms of these variables, the interaction Hamiltonian $H_1$ takes the form of a power series in eccentricities and inclinations, with each term proportional to a periodic function of a linear combination of the orbital angles (except the term proportional to $\cos{(j\lambda_2-j\lambda_1)}$ with $j=0$).  
 
 For systems far from mean motion resonances, these periodic terms are either short period (they average out on timescales comparable to the orbital periods) or secular (varying on long timescales; these terms are independent of the fast angles $\lambda_i$). When the mean motions of the planet form a rational ratio, $n_2/n_1 \sim m/(m+q)$, with $m$ and $q$ positive integers, the system is near a $q-$th order mean motion resonance and all periodic terms in the interaction Hamiltonian which are functions of the combination  $(m+q) \lambda_2-m\lambda_1$ will be slowly varying (relative to the mean longitudes of the planets $\lambda_i$). 
 
 We are interested in the long-term dynamics of two planet systems, and so we will remove all of the short period terms by averaging over the mean longitudes of the planets. There is a canonical transformation between the full Hamiltonian and the averaged Hamiltonian, part of which is derived in the Appendix, which can be important for comparing the analytic and numerical results (see Section \ref{sec:Compare}), though it is not necessary to perform this step explicitly for the following analysis. After averaging, only secular and resonant terms remain in the Hamiltonian. 
 
 The secular terms appear only at second order in the eccentricities and the inclinations or higher. The $q-$th order resonant terms appear at order $q$ in eccentricities and at order $q$ or higher in the inclinations. The first order mean motion resonance terms, with $q=1$, or $n_1/n_2 = P_2/P_1 \sim (m+1)/m$, therefore appear at first order in eccentricities (though they appear at second order in inclinations due to symmetries of the problem). As a result, they are the most important resonances to include when considering resonance overlap of nearly circular and nearly coplanar orbits $-$ other mean motion resonances have smaller widths and contribute less to overlap. Throughout the rest of the paper, an $m$ with no subscript refers to a particular first order resonance near the period ratio $P_2/P_1 \sim (m+1)/m$.  
 
 Truncating the expansion of $H_1$ at first order in $I$ removes all $(I,\Omega)$ dependence (and the secular terms), so that the pair $(G_i,g_i) $ do not appear in the Hamiltonian.  Because the Hamiltonian containing only terms at first order in $e$ or $I$ is the same as the coplanar Hamiltonian at first order in $e$,  any results derived below apply to slightly non$-$coplanar orbits (as long as $I$ is sufficiently small such that $I^2$ terms are negligible).

After performing these steps the resulting Hamiltonian is:
 \begin{align}\label{Ham0}
 H &= H_0(\mathbf{\Lambda})+\epsilon_1 H_1(\mathbf{\Lambda},\mathbf{P},\mathbf{\lambda},\mathbf{p}),
 \end{align}
 where 
 \begin{align}
 H_0 & = -\frac{\mu_1}{2\Lambda_1^2} -\frac{\mu_2}{2\Lambda_2^2},  
 \end{align}
 \begin{align}
 \epsilon_1 & = \frac{m_1}{m_\star},
 \end{align}
 and
 \begin{align}
 H_1 & = -\frac{\mu_2}{\Lambda_2^2}\bigg[ f_{0,1}(\alpha) +\nonumber \\
 & f_{m+1,27}(\alpha)\sqrt{\frac{2P_1}{\Lambda_1}} \cos{((m+1)\lambda_2 -m\lambda_1+p_1)} +\nonumber\\
 &f_{m+1, 31}(\alpha)\sqrt{\frac{2P_2}{\Lambda_2}} \cos{((m+1)\lambda_2 -m\lambda_1+p_2)}  \bigg]
 \end{align}
 and  $\mu_i = G^2 m_\star^2 m_i^3$, $\alpha = a_1/a_2$, $\sqrt{2P/\Lambda} = e(1+O(e^2))$, and $f_{m+1,27}$, $f_{m+1,31}$ and $f_{0,1}$ are functions of Laplace coefficients \citep[p.~539-556]{MurrayDermott}. These can be written as
 \begin{align}
 f_{j,1}(\alpha) & = \frac{1}{2}A_j(\alpha) \nonumber \\ 
 f_{j,27}(\alpha) &= \frac{1}{2}(-2 j-\alpha \frac{d}{d\alpha}) A_j(\alpha) \nonumber \\
 f_{j,31}(\alpha) &= \frac{1}{2}(-1+2j+\alpha \frac{d}{d\alpha}) A_{j-1}(\alpha)\nonumber \\
 A_j(\alpha) & = \frac{1}{\pi} \int_0^{2\pi} \frac{\cos{(j \phi)}}{\sqrt{1-2\alpha \cos{\phi}+\alpha^2}}d \phi.
 \end{align} 
 We could equivalently use $\epsilon_2$ instead of $\epsilon_1$ as our small parameter, the overall coefficient of $H_1$ would just change to $-\mu_1/\Lambda_2^2$. Note that when $m=1$, $f_{2,31}$ must be modified to include a contribution from the indirect terms of $ -2\alpha$.

We approximate these functions as constants, evaluated at the nominal resonance location, $\alpha_{res} \equiv (m/(m+1))^{2/3}$. These values are listed in Table \ref{tab:rvalues} for the first order resonances with $1\leq m \leq13$.
\begin{table}
\centering
  \begin{tabular}{ l | c | c | r }
      \hline
    m &$P_2/P_1$ & $f_{m+1,27}$ & $f_{m+1,31}$ \\ \hline
1 &  2.0&   -1.19049	& 0.42839 \\ \hline 
2 & 1.5 & -2.02522	& 2.48401\\ \hline 
3 & $1.33\bar{3}$ & -2.84043	& 3.28326\\ \hline 
4 & 1.25 & -3.64962	& 4.08371 \\ \hline 
5 & 1.2 & -4.45614	& 4.88471 \\ \hline 
6 & $1.166\bar{6}$ &-5.26125	& 5.68601 \\ \hline 
7 & 1.143 &-6.06552	& 6.48749 \\ \hline 
8 & 1.125 &-6.86925	& 7.28909 \\ \hline 
9 & $1. 111\bar{1}$ &-7.67261	& 8.09077 \\ \hline 
10 &1.1 &-8.47571	& 8.89251 \\ \hline 
11 & $1.09\bar{09}$ &-9.27861	& 9.69429 \\ \hline 
12 & $1.083\bar{3}$ & -10.0814	&  10.4961 \\ \hline 
13 & 1.077 & -10.884	&  11.2979 \\
    \hline
  \end{tabular}
\centering
\caption{The functions $f_{m+1,27}$ and $f_{m+1,31}$ evaluated at the nominal resonance location, with $\alpha = \alpha_{res}(m)$. As $m\longrightarrow \infty$, $f_{m+1,27} \longrightarrow -f_{m+1,31}$.}
 \label{tab:rvalues}
\end{table}
We find numerically that the functions $f_{m+1,27}$ and $f_{m+1,31}$, for $m\geq 2$, evaluated at $\alpha_{res}(m)$ and treated as functions of $m$, are well fit by straight lines. Using a range of $m$ from $2$ to $150$ yields fits of
\begin{align}\label{laplace_fit}
f_{m+1,27} & \approx -0.46-0.802 m \nonumber \\
f_{m+1,31} & \approx 0.87+0.802 m
\end{align}

The maximum fractional deviation in the values for $f_{m+1,27}$ and $f_{m+1,31}$ using the best fit line is $1.9\%$ and $0.4\%$, respectively. It is clear that the ratio of the two quickly approaches $-1$ as $\alpha \longrightarrow 1$.   This is expected for close orbits as in the limit $\alpha \longrightarrow 1$, $f_{27}$ and $f_{31}$ are dominated by the derivative in their expressions \citep{Quillen2011}. Since the coefficient of the derivative contribution is the same (up to a sign) in both expressions, $f_{27} \sim -f_{31}$ for close orbits (high $m$).

 Only one linear combination of $\lambda$'s appears in the Hamiltonian \eqref{Ham0}, implying that we can reduce the number of degrees of freedom by one. Let 
 \begin{align}
 \theta &= (m+1)\lambda_2 -m\lambda_1 \nonumber \\
 \theta_1 &= \lambda_1
 \end{align}
 be the new angles, and $\Theta$ and $\Theta_1$ their corresponding actions.
  The new actions are found using the generating function
 \begin{equation}
 F_2 = \Theta((m+1)\lambda_2 -m\lambda_1) +\Theta_1 \lambda_1
 \end{equation}
  as follows:
  \begin{align}
  \Lambda_1 &= \frac{\partial F_2}{\partial \lambda_1}  = \Theta_1-m\Theta \nonumber \\
  \Lambda_2 &  = \frac{\partial F_2}{\partial \lambda_2}  = (m+1)\Theta,
  \end{align}
  which we can invert as 
  \begin{align}
  \Theta &= \Lambda_2/(m+1) \nonumber \\
  \Theta_1 &= \frac{m}{m+1}\Lambda_2+\Lambda_1.
  \end{align}
  We'll use the canonical polar variables:
  \begin{align} \label{polar}
 x_i &= \sqrt{2P_i}\cos{(p_i)} \nonumber \\
  y_i & = \sqrt{2P_i}\sin{(p_i)},
 \end{align}
 where $x_i$ are momenta and $y_i$ are coordinates ( $\{y_i,x_i\} = 1$, with $\{\cdot,\cdot\}$ denoting a Poisson Bracket).
 This sequence of time independent canonical transformations leads to the Hamiltonian
 \begin{align}\label{Ham1}
 H & = H_0(\Theta_1-m\Theta,(m+1)\Theta) \nonumber \\
 &- \epsilon_1 \frac{\mu_2}{((m+1)\Theta)^2}\bigg[f_{0,1}+\nonumber \\
 &\frac{f_{m+1,27}}{\sqrt{\Theta_1-m\Theta}}(x_1 \cos{\theta} -y_1\sin{\theta}) + \nonumber \\
 &\frac{f_{m+1,31}}{\sqrt{(m+1)\Theta}}(x_2 \cos{\theta} -y_2\sin{\theta})\bigg].
 \end{align}
  
  Note that $\Theta_1$ is an integral of the flow generated by Hamiltonian \eqref{Ham1}.
  
We express all actions in units of $\Theta_1$, and the Hamiltonian in units of $\mu_2/\Theta_1^2$. We choose a new independent variable $\hat{t} = \mu_2/\Theta_1^3 t$, so that when using $\hat{H} = H/(\mu_2/\Theta_1^2) $ as the Hamiltonian function  Hamilton's equations will still hold. Hats denote unitless variables: $\hat{x}_i = x_i/\sqrt{\Theta_1}, \hat{y}_i = y_i/\sqrt{\Theta_1}$, and $\Theta/\Theta_1 = \hat{\Theta}$. We define $\mu_1/\mu_2 = (m_1/m_2)^3 \equiv \zeta^3 $.

 In these units, the Hamiltonian takes the form
 \begin{align}\label{Ham2}
 \hat{H} & = -\frac{\zeta^3}{2(1-m\hat{\Theta})^2}-\frac{1}{2((m+1)\hat{\Theta})^2} \nonumber \\
 & -\epsilon_1 \bigg[\frac{f_{0,1}}{((m+1)\hat{\Theta})^2}+ \nonumber \\
 & (\delta_1 \hat{x}_1 +\delta_2\hat{x}_2) \cos{\theta}-(\delta_1 \hat{y}_1 +\delta_2\hat{y}_2) \sin{\theta}\bigg],
 \end{align}
 where we have defined
 \begin{align}
 \delta_1 & =\frac{1}{((m+1)\hat{\Theta})^2}\frac{f_{m+1,27}}{\sqrt{1-m\hat{\Theta}}} \nonumber \\
 \delta_2 & = \frac{1}{((m+1)\hat{\Theta})^2}\frac{f_{m+1,31}}{\sqrt{(m+1)\hat{\Theta}}} .
 \end{align}

The center of the resonance corresponds approximately to $\hat{\Theta}=\bar{\Theta}$, where $\bar{\Theta}$ is implicitly defined by the expression
\begin{align}
\frac{d \theta}{d \hat{t}}& = \frac{\partial \hat{H}}{\partial \hat{\Theta}} \approx\frac{\partial \hat{H}_0}{\partial \hat{\Theta}}   =0 \nonumber ,
\end{align}
or
\begin{align}
\frac{-m \zeta^3}{(1-m\bar{\Theta})^3}+&\frac{1}{(m+1)^2\bar{\Theta}^3}  = 0,
\end{align}
where we have ignored the order $\epsilon_1$ terms. Including the effects of the $f_{0,1}$ term in solving for $\bar{\Theta}$ is straightforward, but it would only shift the nominal resonance location by order $\epsilon_1$, and we neglect it. Solving for $\bar{\Theta}$ yields
\begin{align}
\bar{\Theta} & = \frac{\alpha_{res}}{m(\alpha_{res}+\zeta)} \nonumber \\
\hat{\Lambda}_{2,res} & = (m+1)\bar{\Theta} \nonumber \\
\hat{\Lambda}_{1,res} & = 1- m\bar{\Theta}.
\end{align}
For orbits near the resonance, $\delta \Theta = \hat{\Theta}-\bar{\Theta}$ is small, and we can expand the Hamiltonian around $\bar{\Theta}$. In this case, the effective Hamiltonian $K = \hat{H} - \hat{H}_0(\bar{\Theta}) +\epsilon_1 f_{0,1}/(m+1)^2/\bar{\Theta}^2$ is 
\begin{align}\label{Ham3}
K & = \frac{\beta}{2} \delta \Theta^2 - \epsilon_1 \times \nonumber\\
&\bigg[(\bar{\delta}_1 \hat{x}_1 +\bar{\delta}_2\hat{x}_2) \cos{\theta}-(\bar{\delta}_1 \hat{y}_1 +\bar{\delta}_2\hat{y}_2) \sin{\theta}\bigg],
\end{align}
where  
\begin{align}
 \bar{\delta}_1 & =\frac{1}{((m+1)\bar{\Theta})^2} \frac{f_{m+1,27}}{\sqrt{1-m\bar{\Theta}}} \nonumber \\
\bar{ \delta}_2 & = \frac{1}{((m+1)\bar{\Theta})^2} \frac{f_{m+1,31}}{\sqrt{(m+1)\bar{\Theta}}} \nonumber \\
\beta  =  \frac{\partial^2 \hat{H}_0}{\partial \hat{\Theta}^2} \bigg|_{\hat{\Theta} = \bar{\Theta}} &= -3 \bigg( \frac{m^2 \zeta^3}{(1-m\bar{\Theta})^4} + \frac{1}{(m+1)^2 \bar{\Theta}^4} \bigg) \nonumber \\
 & = -3 \frac{m^2 (\alpha_{res}+\zeta)^5}{ \zeta \alpha_{res}}.
 \end{align}
 Note that $\delta \Theta$ is the canonical momentum conjugate to $\theta$, and can be rewritten as
\begin{align}\label{deltaTheta}
\delta \Theta & = \frac{\zeta}{m} (1-s)\frac{1}{\alpha_{res}+\zeta+s\cdot \zeta (1+\zeta/\alpha_{res})}
\end{align}
with $s^2 = \alpha/\alpha_{res}$. We see that when $\delta \Theta =0$, or $\Theta = \bar{\Theta}$, $\alpha = \alpha_{res}$.

 We have neglected terms of order $\delta \Theta^3$,   $\epsilon_1 \delta \Theta$, and higher in the Hamiltonian \eqref{Ham3}. If the first two neglected terms are of roughly the same magnitude, this would imply that $\delta \Theta \sim \sqrt{\epsilon_1}$ and that both terms in $K$ are the same magnitude. 
 Note that the expansion about the resonance location is actually $\it{necessary}$ to do - in order for the following sequence of canonical transformations to work out, the functions $\delta_1$ and $\delta_2$ must be constants (we require $\hat{\Theta}$ be replaced by $\bar{\Theta}$).

  Then we define the new canonical set
 \begin{align}\label{trans}
 r_1 & = \frac{\bar{\delta}_1 \hat{x}_1 +\bar{\delta}_2 \hat{x}_2}{\bar{\delta}} \nonumber \\
 s_1 & = \frac{\bar{\delta}_1 \hat{y}_1 +\bar{\delta}_2 \hat{y}_2 }{\bar{\delta}} \nonumber \\
 r_2 & = \frac{\bar{\delta}_2 \hat{x}_1 -\bar{\delta}_1 \hat{x}_2}{\bar{\delta}} \nonumber \\
 s_2 & = \frac{\bar{\delta}_2 \hat{y}_1 -\bar{\delta}_1 \hat{y}_2 }{\bar{\delta}} .
 \end{align}
 The $r_i$ are momenta, and $s_i$ their canonical conjugate coordinates, and  $\bar{\delta} = \sqrt{\bar{\delta}_1^2 + \bar{\delta}_2^2}$.  Substituting Equation \eqref{trans} into the Hamiltonian \eqref{Ham3} yields:
 \begin{align}\label{Ham4}
K & = \frac{\beta}{2} \delta \Theta^2 - \epsilon_1 \bar{\delta} \times \bigg[r_1 \cos{\theta}-s_1\sin{\theta}\bigg]
\end{align}
Note that $r_2$ and $s_2$ are conserved.
Next, let $r_1 = \sqrt{2 \Psi} \cos{\psi}$ and $s_1 =  \sqrt{2 \Psi} \sin{\psi}$.  $\Psi$ is proportional to $e^2$:
\begin{align}\label{definePsi}
\Psi & = \frac{1}{2}(r_1^2+s_1^2) \nonumber \\
& = \frac{\zeta}{\alpha_{res}+\zeta}\frac{1}{2 (R^2 +\zeta \sqrt{\alpha_{res}})}  \times \nonumber \\
& \bigg[ R^2 e_1^2 + e_2^2 - 2 R e_1 e_2 \cos{(\Delta \varpi)} \bigg] \nonumber \\
& = g_m R^2 \bigg[ e_1^2 + e_2^2/R^2 - 2/R e_1 e_2 \cos{(\Delta \varpi)} \bigg]
\end{align}
with $R = \abs{f_{m+1,27}/f_{m+1,31}}$  and $g_m$ implicitly defined. If we define: $\mathbf{e}_1 \equiv e_1 (\cos{\varpi_1}, \sin{\varpi_1})$ and the same for $\mathbf{e}_2$, $\Psi \propto ( \mathbf{e}_1-  \frac{1}{R} \mathbf{e}_2)^2$, so it is a measure of the total eccentricity, and we will refer to the quantity 
\begin{align}\label{definesigma}
\sigma &\equiv \sqrt{ e_1^2 + e_2^2/R^2 - 2/R e_1 e_2 \cos{(\Delta \varpi)}} = \bigg[\frac{\Psi}{g_m R^2}\bigg]^{1/2}
\end{align}
as the weighted eccentricity of the system. The angle $\psi=\arctan{\{ s_1/r_1\}}$ is a generalized longitude of pericenter. In the special case when $e_2=0$, $\psi = \pi-\varpi_1$, and when $e_1=0$, $\psi = -\varpi_2$. These new ``eccentricity" vector components $(r_i, s_i)$ are obtained from a linear transformation of the original $(x_i,y_i)$. As pointed out by \cite{Batygin}, the transformation is a rotation - it preserves length, and indeed you can show that  $\Psi +\Psi_2 = (P_1+P_2)/\Theta_1$, where $\Psi = \frac{1}{2}(r_1^2+s_1^2)$ and $\Psi_2 = \frac{1}{2}(r_2^2+s_2^2)$. Therefore $\Psi+\Psi_2$ is proportional to the angular momentum deficit.

In terms of $\Psi$ and $\psi$, the Hamiltonian $K$ becomes
  \begin{align}\label{Ham5}
K & = \frac{\beta}{2} \delta \Theta^2 - \epsilon_1 \bar{\delta} \sqrt{2 \Psi} \cos{(\theta+\psi)}.
\end{align}
And now we will choose new angles $\phi = \theta+\psi$ and $\gamma = -\theta$. Using the generating function 
\begin{equation}
F_2 = \Phi(\theta+\psi) - \Gamma \theta,
\end{equation}
the new set of actions is
\begin{align}
\Psi &= \Phi \nonumber \\
\delta \Theta &= -\Gamma+\Phi.
\end{align}
Then the Hamiltonian becomes
  \begin{align}\label{Ham6}
K & = \frac{1}{2}{\beta} (\Phi-\Gamma)^2 - \epsilon_1 \bar{\delta} \sqrt{2 \Phi} \cos{\phi},
\end{align}
with $\Gamma$ an integral of the motion and $\bar{\Theta}$ a constant. At this point, we are essentially finished. The three conserved quantities $\Gamma$, $\Theta_1$, and $r_2$ make the original four degree of freedom problem a one degree of freedom problem\footnote{ The conserved quantity $s_2$ is not in involution with $r_2$ as $\{s_2,r_2\} = 1$, so there are only three independent conserved integrals of the motion.}. The total angular momentum is a function of these conserved quantities.  However, in order to simplify the analysis, it is worthwhile to do a bit more algebra in order to reduce the Hamiltonian to a one parameter function.

To that end, we rescale the variables. Let actions be divided by the unitless parameter $Q$; angles are unchanged. Let primes mark the rescaled variables. We also rescale both the Hamiltonian $K$ and the time $t$ as $H' = K/a$ and $t' = \hat{t}/a$. After this step, the Hamiltonian is
\begin{align}
H' & = K/a \nonumber\\
&= -\frac{1}{2}|\beta| Q^2 \frac{1}{a}(\Phi'-\Gamma')^2 -\epsilon_1 \bar{\delta} \sqrt{Q} \frac{1}{a}  \sqrt{2 \Phi'} \cos{\phi}.
\end{align}

If we choose $a = Q^2 |\beta|$, and $Q = (\epsilon_1 \bar{\delta}/|\beta|)^{2/3}$, we are only left with one free parameter ($\Gamma'$). It can be shown that

\begin{align}
Q & = \epsilon_p^{2/3} \zeta \bigg(\frac{f_{m+1,31}^2}{9(m+1)(1+\zeta)^2}\bigg)^{1/3}\frac{\alpha_{res}^{5/6}}{m} \bigg(\frac{R^2+\zeta\sqrt{\alpha_{res}}}{(\alpha_{res}+\zeta)^5}\bigg)^{1/3},
\end{align}
where we have replaced $\epsilon_1 = \zeta \epsilon_p/(1+\zeta)$ with the total mass of the planets $\epsilon_p=\epsilon_1+\epsilon_2$.

This rescaling results in a Hamiltonian function
\begin{align}
H' & = -\frac{1}{2} (\Phi'-\Gamma')^2 - \sqrt{2 \Phi'} \cos{\phi}.
\end{align}
We perform a final canonical transformation
\begin{align}
X &= \sqrt{2 \Phi'} \cos{\phi} \nonumber \\
Y & = \sqrt{2 \Phi'} \sin{\phi} ,
\end{align}
so that - at last!
  \begin{align}\label{Ham7}
H' & = -\frac{1}{2} \bigg[ \frac{1}{2}(X^2+Y^2) -\Gamma' \bigg]^2 -X
\end{align}
This is the one degree of freedom Hamiltonian with a single free parameter that we have been seeking. It is interesting that the conserved quantities $s_2$ and $r_2$ do not appear at all, not even as constant parameters.  We note that the Hamiltonian \eqref{Ham7}, which applies for arbitrary mass ratio $\zeta$, has the same functional form as the Hamiltonian for the motion of a test particle near a first order resonance with a planet ($\zeta =0$ or $\zeta \longrightarrow \infty$). It must be true that in either limit the Hamiltonian \eqref{Ham7}  reduces to the restricted case. 

However, we have found something stronger to be true: the Hamiltonian \eqref{Ham7} is approximately independent of the mass ratio $\zeta$. To show this, we write $s = \sqrt{\alpha/\alpha_{res}}\approx1+\Delta \alpha/(2\alpha_{res})$, and expand Equation \eqref{deltaTheta} in powers of $\Delta \alpha/\alpha_{res}$, 
\begin{align}\label{usefulLater}
\delta \Theta & = -\frac{\alpha_{res} \zeta}{m(\alpha_{res}+\zeta)^2}\frac{1}{2}\frac{\Delta \alpha}{\alpha_{res}}+\ldots
\end{align} 
so that
\begin{align}
\Gamma' &= \frac{\Psi - \delta \Theta}{Q} \nonumber \\
& = \frac{1}{Q} \times \bigg(g_m  R^2\sigma^2  +\frac{\zeta}{m(\alpha_{res}+\zeta)^2}\frac{1}{2}\Delta \alpha\bigg)\nonumber \\
& = \frac{1}{2}\bigg(\frac{9 (m+1)}{\epsilon_p^2 f_{m+1,31}^2\alpha_{res}^{5/2}}\bigg)^{1/3}\bigg(\frac{(1+\zeta)^2(\alpha_{res}+\zeta)^2}{(R^2+\zeta\sqrt{\alpha_{res}})} \bigg)^{1/3} \times \nonumber \\
& \bigg[ \frac{ mR^2\sigma^2}{R^2+\zeta \sqrt{\alpha_{res}}}+ \frac{\Delta \alpha}{\alpha_{res}+\zeta}\bigg].
\end{align}

We are considering close orbits, where $\alpha_{res} \approx 1, R^2 \approx 1$, and $\sigma \approx \sqrt{e_1^2 +e_2^2 -2 e_1 e_2 \cos{\Delta \varpi}}$. In this limit 
\begin{align}\label{nomads}
\Gamma' & \sim \frac{1}{2}\bigg(\frac{9 (m+1)}{\epsilon_p^2 f_{m+1,31}^2}\bigg)^{1/3}\bigg(\frac{(1+\zeta)^2(1+\zeta)^2}{(1+\zeta)} \bigg)^{1/3} \times \nonumber \\
& \bigg[ \frac{m  \sigma^2}{1+\zeta}+\Delta \alpha \frac{1}{1+\zeta}\bigg] \nonumber \\
& = \frac{1}{2}\bigg(\frac{9 (m+1)}{\epsilon_p^2 f_{m+1,31}^2}\bigg)^{1/3} \times  \bigg[ m  \sigma^2 +\Delta \alpha \bigg] .
\end{align}

The mass ratio dependence of the Hamiltonian drops out for close orbits near first order mean motion resonances. Although the function form of the Hamiltonian \eqref{Ham7} is identical to that for a test particle and a planet near a first order mean motion resonance, we would not have a priori expected the coefficients to work out in such a way. Note that this does \textit{not} imply that the actual motion is independent of mass ratio - for example, the relative amplitude of the eccentricity oscillation of the two planets is still proportional to the mass ratio of the planets. 

Also, the mass ratio will certainly be a factor for the 2:1 mean motion resonance, where $\alpha$ is further from unity and $R = 2.78$ (and hence the approximations used to get to expression \eqref{nomads} do not hold). The significant deviation of $R$ from $1$ is because of the indirect contribution to the disturbing function for this commensurability. 

\subsection{Fixed Point Analysis and Conditions for Resonance}
\begin{figure}
	\begin{center}
	\includegraphics[width=3.5in]{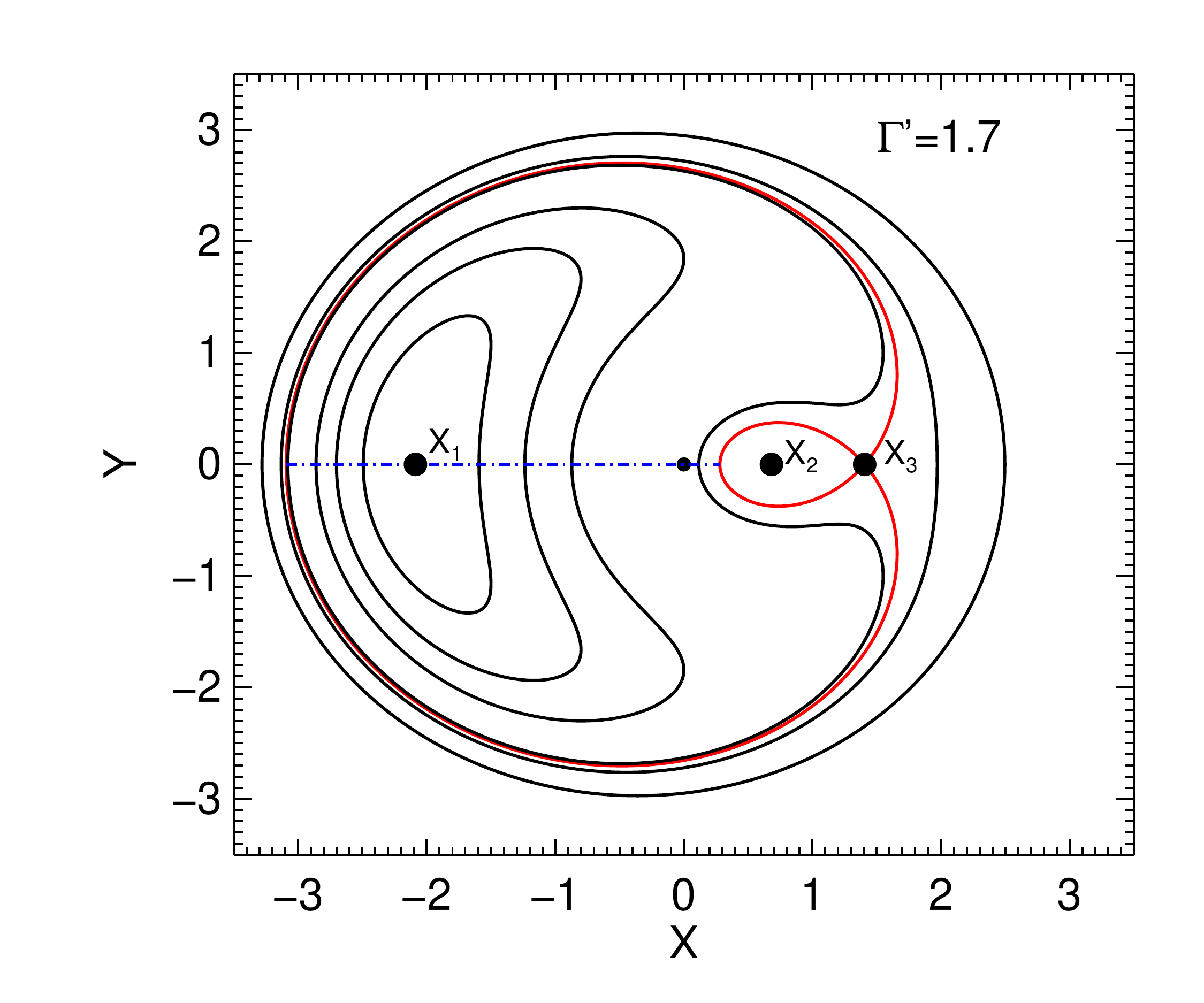}
	\caption{Contours of the Hamiltonian function \eqref{Ham7} for a value of $\Gamma'=1.7$. The separatrix is marked in red. The fixed point with the largest value of $X$, $X_3$, is the unstable fixed point, whereas the other two are centers. Note that some resonant orbits correspond to circulation of the resonant angle (resonant contours which enclose the origin, which is marked with a small black dot). It is only when $\Gamma' \geq 1.88988$ that the seperatrix encloses only librating orbits.  The dashed blue line illustrates the resonance width.}
	\label{fig:contour}
	\end{center}
	\end{figure}

The analysis of the level curves and fixed points of the Hamiltonian \eqref{Ham7} is the same as in the case of the second fundamental model for resonance (also called an Andoyer Hamiltonian with index 1, see for example \citet{HenrardAgain, FerrazBook}). To summarize, the fixed points satisfy the equations:
\begin{align}\label{FP}
\frac{d X}{d {t}'}  = 0 & \longrightarrow Y\bigg(\frac{1}{2}(X^2+Y^2)-\Gamma'\bigg) = 0 \nonumber \\
&\longrightarrow Y=0 \nonumber \\
\frac{d Y}{d {t}'}  = 0 &\longrightarrow X\bigg(\frac{1}{2}(X^2+Y^2)-\Gamma'\bigg) +1 = 0 \nonumber \\
&\longrightarrow X^3-2\Gamma'X+2 =0 .
\end{align}
When $Y=0$ and $X>0$, the fixed point corresponds to $\phi=0$.  When $Y=0$ and $X<0$, the fixed point corresponds to $\phi=\pi$. 
The discriminant for the cubic equation \eqref{FP} is $\Delta = 32 (\Gamma'^3-27/8)$, but only the sign of $\Delta$ matters for the number of real roots. If $\Delta > 0$, there are three real roots; when $\Delta<0$, there is only a single real root. This is the only bifurcation for this system. The transition corresponds to $\Gamma'=3/2$.  When there are three real roots, the root with the largest value of $X$ is a saddle point, and the other two are centers (elliptic fixed points). Two homoclinic orbits, together comprising the separatrix, are the level curve connecting the unstable fixed point $X_3$ to itself, as shown in Figure  
\ref{fig:contour} in red. When the unstable fixed point and the stable fixed point at $X_2$ merge at the value of $\Gamma' = 3/2 $, a single center remains (at $X_1$), and there is no longer a separatrix. This Hamiltonian is only a function of $Y^2$, so there is symmetry in the contours of $H(X,Y)$ about the $Y=0$ line.

We clarify here the distinction between an orbit with an oscillating resonant angle and an orbit ``in resonance". An orbit in resonance evolves within the resonant region, which is the area located between the two homoclinic orbits, corresponding to oscillations around $X_1$ \citep{HenrardAgain, FerrazBook}. But depending on the value of $\Gamma'$, not all of the enclosed resonant orbits correspond to oscillatory motion of $\phi$, and there also exists oscillatory motion of $\phi$ outside of the resonant region (and also when there is no separatrix at all). Whether or not the resonant angle circulates or oscillates depends on the choice of the origin; the intrinsic dynamics do not depend on coordinate choice.

\begin{figure}
	\begin{center}
	\includegraphics[width=3.5in]{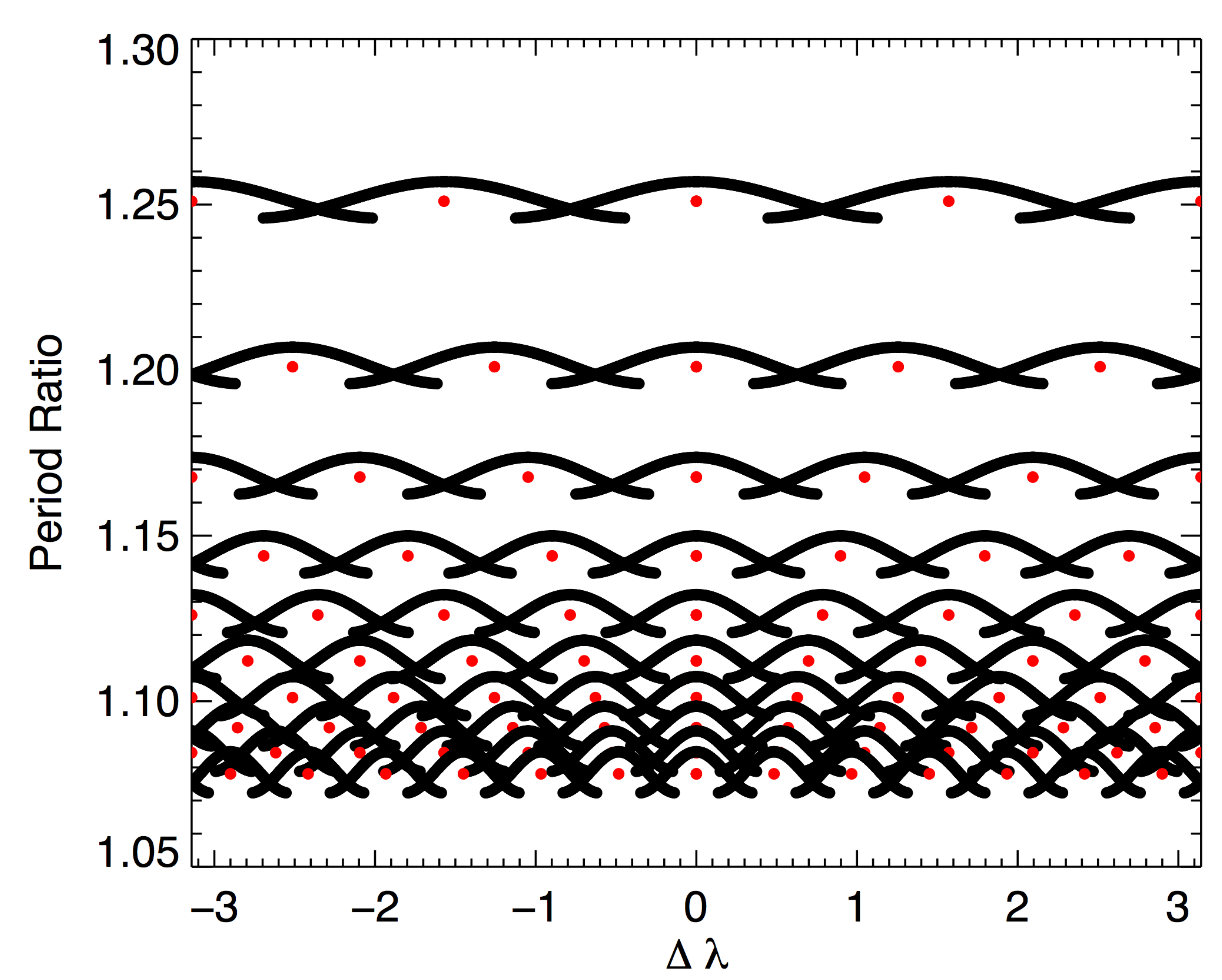}
	\caption{Seperatrices of the first order mean motion resonances at zero eccentricity as a function of period ratio and initial $\Delta \lambda$. All angles but $\lambda_1$ are set to zero, so using Equation \eqref{formula}, $\Delta \lambda = (2k+1)\pi/m$ corresponds to $\phi=\pi$. There are $m$ separate resonance centers, marked in red, in $\Delta \lambda$ at the $m$:$m+1$ resonance. Note that the separatrices do not close on one side because one of the boundaries of the separatrix does not extend through all values of $\phi$. The masses of the planets are set to $\epsilon_1=\epsilon_2 = 10^{-5}$. At period ratios close to one, the seperatrices associated with first order resonances of different $m$ overlap.}
	\label{fig:ChainSep}
	\end{center}
	\end{figure}

The fixed point $X_1$ at the center of the resonance region corresponds to $\phi=\pi$, or 
\begin{align}\label{formula}
\lambda_2 + m \Delta \lambda + \psi &= (2 k+1) \pi \nonumber \\
\Delta \lambda &= \frac{(2k+1)\pi-\psi -\lambda_2}{m},
\end{align}
where $k$ is an integer. This implies that there should be $m$ values of $\Delta \lambda$ corresponding to a single center in the $(X,Y)$. In terms of the original angles, there are $m$ resonant islands for the $m$:$m+1$ resonance. This is exhibited Figure \ref{fig:ChainSep} for initially zero eccentricity orbits.

\section{RESONANCE OVERLAP}\label{sec:Resoverlap}
The motion determined by an integrable $2n$ degree of freedom Hamiltonian reduces to that of $n$ angles $\phi_i$ changing at constant rates $\omega_i$. When the unperturbed frequencies $\omega_i$ are commensurate, satisfying $\sum_{i=0}^{n-1} a_i \omega_i \approx 0$, where $a_i$ is an integer and at least two of the $a_i$ are nonzero, the system is near a resonance (with  $\sum_{i=0}^{n-1} a_i = \phi_a$ as the resonant angle). When the system is weakly perturbed (in such a way as to be non-integrable), if only one such integer vector $\mathbf{a} =( a_0,Éa_{n-1})$ exists, the evolution of orbits near the resonance is approximately integrable. However, if there exists another (independent) integer vector $\mathbf{b}$ such that $\sum_{i=0}^{n-1} b_i \omega_i \approx 0$, with a corresponding resonant angle $\phi_b$, the motion is more complicated. The resonance overlap criterion states that for a weakly perturbed (and non-integrable) Hamiltonian system, if more than one linearly independent combination of the angles, treated individually and without interaction, are simultaneously resonant, the motion will be chaotic.  The region in which a particular angle is resonant is within the boundaries of the seperatrix, and so we must first determine the widths of the first order resonances, in terms of their seperatrices, in order to determine if resonance overlap occurs.

 The resonance overlap criterion has been applied to the first order mean motion resonances in the case of an eccentric test particle and a circular planet orbiting a much more massive central body \citep{Wisdom1980}. It was found that the orbit of a test particle with initially zero eccentricity will be chaotic if its semi major axis $a$ satisfies $\abs{a-a_{pl}}/a_{pl}\lesssim1.3 \epsilon_{pl}^{2/7} $ (where the subscript $pl$ stands for planet) unless the test particle is protected by the 1:1 resonance. An alternative analytic derivation confirmed this, though accompanying numerical integrations indicated that the coefficient is slightly larger ($\sim 1.5$, \citet{DuncanQuinnTremaine}). This result was extended to slightly eccentric particle orbits by \citet{MustillWyatt}, who found the overlap region satisfied $\abs{a-a_{pl}}/a_{pl} \lesssim 1.8 (\epsilon_{pl} e)^{1/5}$ when $e \gtrsim 0.2 \epsilon_{pl}^{3/7}$. We are interested here in determining how these thresholds translate in the case of two massive planets on eccentric orbits. Our approach closely follows that of \citet{Wisdom1980}.

\subsection{Determining the Widths of the Resonances}
Our goal is to determine the boundary of the resonant region, defined by the separatrix, as a function of orbital elements. The resonant region only appears when $\Gamma'>3/2$, so we only consider this case. For a given value of $\Gamma' > 3/2$, we first determine the position of the unstable fixed point $X_3$. The value of the Hamiltonian along the separatrix is $H'(X_3,0) = H_{sep}$. We will quantify  the ``width" of the resonant region as the maximal distance between the inner and outer curve of the separatrix. This maximal width is the distance in $X$ between the two crossings of the separatrix with the $X$ axis, marked with a blue dotted line in Figure \ref{fig:contour}. The crossing points are the solutions of the equation $H(X,0) -H_{sep}=0$ (where one of the roots is $X_3$).  

It can be shown (see for example \citet{FerrazBook}) that the crossing points $X_{\star 1 }$ and $X_{\star 2 }$ have values of
\begin{align}\label{crossings}
X_{\star 1 } & = -X_3+2/\sqrt{X_3} \nonumber \\
X_{\star 2 } & = -X_3-2/\sqrt{X_3} ,
\end{align}
and that 
\begin{align}\label{diff}
\Delta X = X_{\star 1}-X_{\star 2} &= 4/\sqrt{X_3}.
\end{align}

This width in terms of $X$ is related to the width in terms of $\Psi'$ and $\delta \Theta'$ as
\begin{align}\label{widthformula}
\Delta \Psi' &= \Psi'_{\star 1} - \Psi'_{\star 2} \nonumber \\
& =\frac{1}{2}(X_{\star 1}-X_{\star 2})\abs{X_{\star 1} + X_{\star 2}} = 4 \sqrt{X_3} \nonumber \\
\Delta ( \delta \Theta' )& = \Delta \Psi',
\end{align}
where we have used the conserved quantity  $\Gamma' = \Psi' - \delta \Theta'$. The boundaries of the resonance in terms of the scaled quantities $(\delta \Theta',\Psi')$ are the same for every resonance (but there is dependence on $m$ contained in the scaling factor Q).

The boundaries in terms of  $\Psi'$ and $\delta \Theta'$ correspond uniquely to a width in terms of the weighted eccentricity $\sigma$ and the semimajor axis ratio of the planets, through Equations \eqref{deltaTheta} and \eqref{definePsi}.  The weighted eccentricity is a weak function of the semimajor axis ratio, as $R$ depends on the Laplace coefficients.

\begin{figure}
	\begin{center}
	\includegraphics[width=3.5in]{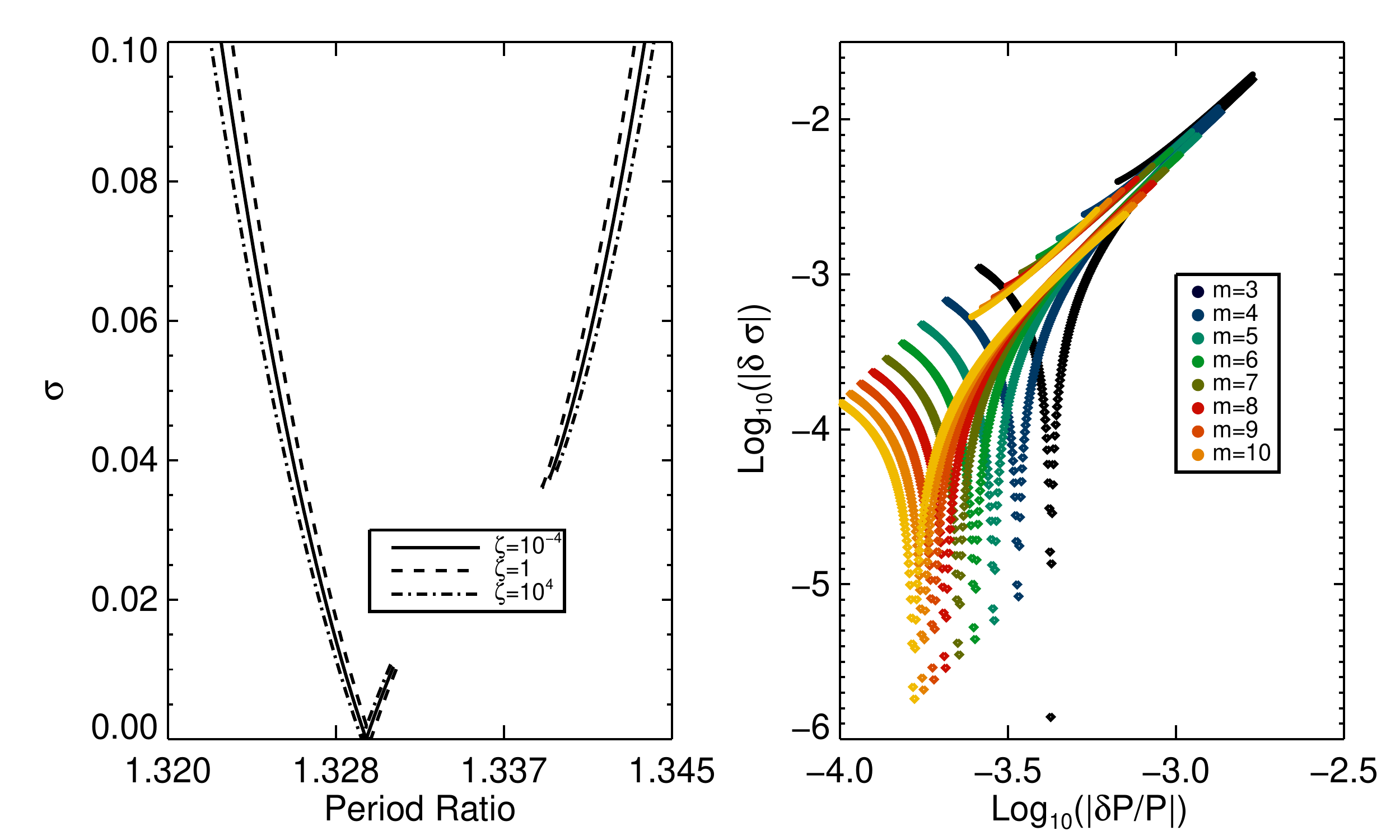}
	\caption{\textit{Left:} Boundaries of the $m=3,$ 4:3 mean motion resonance as a function of the period ratio of the planets and their weighted eccentricity $\sigma$ (Equation \ref{definesigma}). The two sets of dashed lines correspond to cases of an interior or exterior test particle with a massive planet, while the solid line shows the resonance bounds for the case of equal mass planets. In all cases the total mass of the planets is $2\times 10^{-5}$ relative to mass of the central body. \textit{Right:} The difference in period ratio and $\sigma$ of the resonance boundaries between the case of $\zeta = 10^{-4}$ and $\zeta = 10^4$. Plotted is $\log_{10}{\abs{P_2/P_1(\zeta=10^4)-P_2/P_1(\zeta=10^{-4}))}/[(m+1)/m]}$, i.e. the difference in period ratio between the two cases, relative to the period ratio at the nominal resonance location, and $\log_{10}{[\sigma(\zeta=10^4)-\sigma (\zeta=10^{-4})]}$ for several of the first order resonances. Although $\zeta$ changes by 8 orders of magnitude, the resonance widths only change by a small amount.}
	\label{fig:MRtwo}
	\end{center}
	\end{figure}

The resulting boundaries for a single first order resonance are shown in the period ratio - $\sigma$ plane in Figure \ref{fig:MRtwo}. To make this plot, only $\epsilon_1$, $\epsilon_2$, and the particular resonance integer $m$ are required. Three instances of the resonance boundaries are plotted corresponding to the cases of $\zeta =10^{-4}, \zeta = 1$, and $\zeta =10^4$, confirming that the mass ratio of the planets  $\zeta$ is relatively unimportant for determining resonance widths (and therefore that we only require $\epsilon_p$, not both $\epsilon_1$ and $\epsilon_2$). Note that orbits inside the resonance have different libration amplitudes about the fixed point and do not in general explore the full width of the resonance.  Since $\Psi'_{\star 1} = X_{ \star  1}^2/2$ passes through zero, $X_{\star 1}$ corresponds to the left boundary of the seperatrix (at smaller period ratios) in Figure \ref{fig:MRtwo}.

\subsection{A Resonance Overlap Criterion for Initially Circular Orbits}\label{subsec:ResOver}
We now turn to the question of resonance overlap. If the overlap of two resonances is moderate, only the high amplitude libration orbits (oscillating around the resonant fixed point) will be chaotic (since only the high amplitude libration orbits come very near the seperatrix). For a particular value of $\sigma$, however, if the overlap is extreme enough that even the resonance centers are in the overlapped region, all resonant motion at higher values of $\sigma$ will be chaotic.  If the resonances overlap so much that even initially zero eccentricity orbits are chaotic, then approximately all orbits with higher eccentricities will be chaotic as well. As a result, the period ratio at which the first order resonances overlap, even for initially zero eccentricity orbits, is a critical period ratio. Essentially all systems of planets with orbits more tightly packed than this critical period ratio will be chaotic, regardless of the eccentricities.

The seperatrix can only pass through zero eccentricity when $X_{\star 1}=0$. Using Equation \eqref{crossings}, this implies that $X_3 = 4^{1/3}$. Therefore the width of the resonance at zero eccentricity is $\Delta ( \delta \Theta' ) = 4 \sqrt{X_3} = 5.04$. Since we are working in scaled variables, this width is the same for all $m$. If we substitute this value of $X_3$ into Equation \eqref{FP}, we find that the value of $\Gamma'$ corresponding to zero eccentricity on the seperatrix is $\Gamma' = 3/4^{1/3} = 1.88988$. Since $\Psi' = 0$ at this point, $\delta \Theta' = -1.88988$ at zero eccentricity on the seperatrix. In the original application of the first order resonance overlap criterion to the restricted three body problem, Wisdom defined an effective resonance width as symmetric about the resonance center, i.e. as $\Delta (\delta \Theta') = 1.88988-(-1.88988) = 3.78$. This is a very slight underestimate.

A width in terms of $\delta \Theta'$ is related to a width in terms of $\alpha$ by solving for $s$ in Equation \eqref{deltaTheta},
 \begin{align} \label{solve}
 s =\sqrt{\frac{\alpha}{\alpha_{res}}}& = \frac{1-\delta \Theta' Q m/(\alpha_{res}/\zeta+1)}{1+\delta \Theta' Q m/(1+\zeta/\alpha_{res})}.
 \end{align}

We Taylor expand the expression for $s$ in powers of $\epsilon_p$. The result is
\begin{align}
s^2 & \sim 1-\delta \Theta' c_0 \epsilon_p^{2/3} + O(\epsilon_p^{4/3})) \nonumber \\
c_0 & = 2  \bigg[\frac{ f_{31}^2 (R^2 +\sqrt{\alpha_{res}} \zeta) (\alpha_{res}+\zeta)}{ 9 \sqrt{\alpha_{res}}(m+1)(1+\zeta)^2} \bigg]^{1/3} .
\end{align}

Therefore the width of the resonance $\delta \alpha_m/\alpha_{res} $ is 
\begin{align}
\frac{\delta \alpha_m}{\alpha_{res}}  &= s^2(\delta \Theta'_1) -s^2(\delta \Theta'_2) \nonumber \\
 &=  \Delta (\delta \Theta') \epsilon_p^{2/3}c_0 + O(\epsilon_p^{4/3}) .
\end{align}
Note that if the resonance bounds were symmetric there is no contribution at $O(\epsilon_p^{4/3})$, since that term depends on $(\delta \Theta^{'2}_2 - \delta \Theta^{'2}_1)$.

When overlap occurs, $\alpha$ is near unity, so we will replace $R^2$ with 1 and $\alpha_{res}$ with 1. Moreover, we'll replace $f_{m+1,31}$ with $0.802 m+0.87 \sim 0.802 m$ for large $m$. This approximation assumes $1/m$ is small. We will also neglect the $\epsilon_p^{4/3}$ terms in this approximation, and will justify it afterwards. 
Then the width of the resonance for close orbits (high $m$) is
\begin{align}\label{instep}
\frac{\delta \alpha}{\alpha_{res}}  &=\frac{2}{9^{1/3}}\Delta (\delta \Theta')  \epsilon_p^{2/3}\bigg[\frac{ (0.802m)^2 (1 +\zeta) (1+\zeta)}{ (m+1)(1+\zeta)^2} \bigg]^{1/3} \nonumber \\
& \approx 2\bigg(\frac{0.802}{3}\bigg)^{2/3}\Delta (\delta \Theta') \epsilon_p^{2/3}\bigg[\frac{ m^2}{ (m+1)} \bigg]^{1/3} \nonumber \\
& \approx 4.18 \epsilon_p^{2/3} m^{1/3}.
\end{align}
where in the last line we have used $\Delta (\delta \Theta')=5.04$.  The mass ratio $\zeta$ can only appear in the higher order terms, confirming the weak dependence of the widths on the mass ratio. 

The distance between two neighboring resonances is 
\begin{align}
\Delta \alpha & = \frac{d}{dm}\alpha_{res}(m) \Delta m \nonumber \\
& = \frac{d}{d m} \bigg(\frac{m}{m+1}\bigg)^{2/3} (1) \nonumber \\
& = \alpha_{res}(m) \frac{2}{3 m^2} \frac{m}{m+1} \nonumber \\
& \sim \alpha_{res}(m)   \frac{2}{3 m^2}  \nonumber \\
\frac{\Delta \alpha}{\alpha_{res}}  &\sim  \frac{2}{3 m^2}.
\end{align}

When the distance between two neighboring resonances is equal to the sum of their half widths, the resonance overlap criterion is satisfied. This occurs when
\begin{align}\label{simple1}
\frac{\Delta \alpha}{\alpha_{res,m}}  &\approx 0.5\times \bigg(\frac{\delta \alpha}{\alpha_{res,m }}+\frac{\delta \alpha}{\alpha_{res,m+1}} \bigg)\nonumber \\
\frac{2}{3 m^2} & \sim  4.18  \epsilon_p^{2/3} m^{1/3} \nonumber \\
m & \sim 0.455 \epsilon_p^{-2/7} .
\end{align}
Hence all orbits should be chaotic if their $\mathbf{averaged}$ period ratio satisfies
\begin{align}\label{overlapcrit2}
\frac{P_2}{P_1} &\lesssim 1+\frac{1}{m} \nonumber \\
&\lesssim 1+ 2.2\epsilon_p^{2/7},
\end{align}
or equivalently
\begin{align}\label{crit_overlap}
\frac{a_2-a_1}{a_1}\lesssim 1.46 \epsilon_p^{2/7},
\end{align}
and the planets are not protected by the 1:1 resonance. We note that this criterion is for averaged coordinates, but in most cases the transformation between the averaged and true Hamiltonian is negligible.  
 This also must hold for the restricted case; indeed, this result has the same function form as that derived by \citet{Wisdom1980}, albeit with a $\sim10\%$ larger coefficient. The numerical coefficient of $1.46$ is dependent on our specific definition of the width of the resonance. If we had carried through the calculation using $\Delta (\delta \Theta') = 3.78$, as Wisdom did, in Equation \eqref{instep}, we would have found  $m\sim 0.51  \epsilon_p^{-2/7}$, or $(a_2-a_1)/a_1 \lesssim 1.3 \epsilon_p^{2/7}$ in agreement with the \citet{Wisdom1980} result. 

This criterion should be interpreted as a minimum criterion for widespread chaos. We have neglected, for example, the effect of higher order resonances. The chaotic separatrices of these resonances serve to link two neighboring first order resonances before they would overlap if there were not higher order resonances present. We have also ignored the finite extent of the chaotic zone around the seperatrix, though this effect is probably less important than the first.

 In the above calculation, we kept terms of order $1/m^2$ and $\epsilon_p^{2/3} m^{1/3}$, and we neglected terms of order $\epsilon_p^{2/3}/m^{2/3}, 1/m^3$, and $\epsilon_p^{4/3} m^{1/3}$.  For Jupiter mass planets, $\epsilon_p \sim 10^{-3}$, and the criterion predicts overlap at $m\sim3$, and the neglected terms are of order $5\times10^{-3}, 3\times10^{-2}$, and $10^{-4}$, respectively. For smaller mass planets, the error incurred by neglecting the $\epsilon_p^{4/3}$ terms will be smaller than the that incurred by neglecting $\epsilon_p^{2/3}/m^{2/3}$ and $1/m^3$ terms. This justifies ignoring the $\epsilon_p^{4/3}$ terms in the above calculation. However, it is clear that the approximation of $\alpha_{res} =1$, $R^2 =1$, and $1/m \ll 1$ become less accurate for high mass planets in the regime where complete resonance overlap occurs. We will study these effects numerically.

\subsection{A Resonance Overlap Criterion for Initially Eccentric Orbits}\label{sec:resEcc}
\begin{figure}
	\begin{center}
	\includegraphics[width=3.5in,keepaspectratio=true]{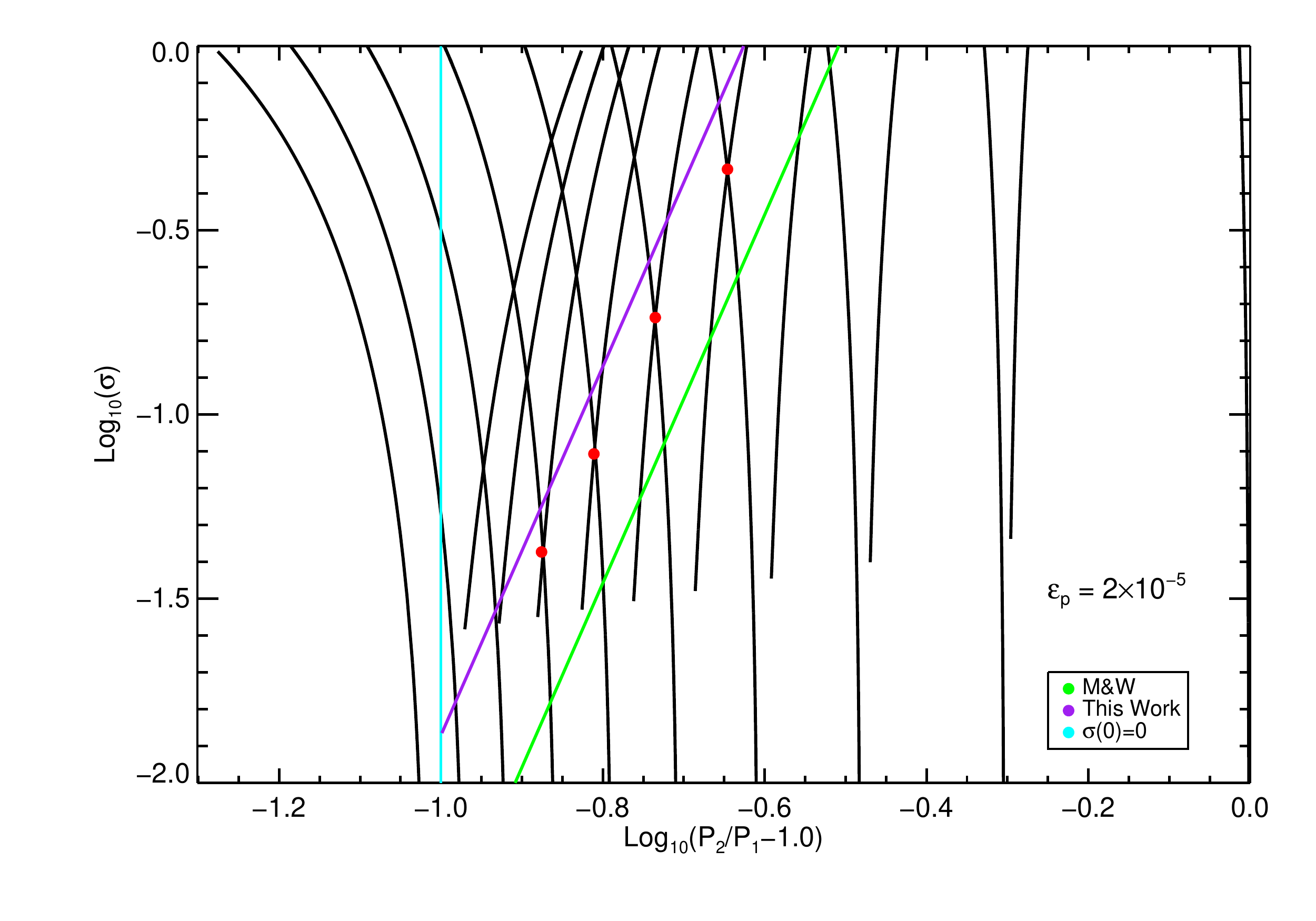}
	\caption{The separatrices of the first order mean motion resonances for arbitrary eccentricities (in black) when $\epsilon_p = 2\times10^{-5}$. Note that although the seperatrices are plotted for all $\sigma<1$, they do not apply in reality to high eccentricity systems! However, they still can be used to test an overlap criterion as a function of $\sigma$. The positions where neighboring first order resonances first overlap are marked by red dots. In both plots, we show the resonance overlap criterion for initially circular orbits (in purple), the estimate derived here for arbitrary eccentricites (in blue), and the \citet{MustillWyatt} criterion (in green). The curves are an approximation to the red dots. }
	\label{fig:diagram}
	\end{center}
	\end{figure}

We now briefly address developing a resonance overlap criterion as a function of eccentricity (while still assuming eccentricities are small, so that our approximation of the Hamiltonian holds). The resonant widths grow with eccentricity, and so we expect resonance overlap to occur at a period farther from unity when eccentricities are nonzero. Figure \ref{fig:diagram} shows the separatrices of several first order mean motion resonances in the case of $\epsilon_p = 2\times10^{-5}$; a resonance overlap criterion for initially eccentric orbits would pass through the crossing points (shown in red) of neighboring resonances. We remind the reader that the the theory does not apply for real systems with large eccentricities; however any analytic overlap criterion derived from the theory should be able to predict where overlap occurs even in a regime where the theory does not apply, and so we consider large eccentricities here as an exercise. 

An approximate resonance overlap criterion as a function of eccentricity has been obtained for the restricted three body problem \citep{MustillWyatt}, and so we expect to be able to recover the results obtained in that case. To reproduce their result, we look at the large $\Gamma'$ limit (which we will show amounts to a large eccentricity limit). 

As $\Gamma'$ increases, the value of the unstable fixed point $X_3$ grows. The equation for the fixed points \eqref{FP} in this case reduces to $X^3-2\Gamma'X +2 \approx X^3-2\Gamma'X =0$. This admits $X=0$ and $X=\pm \sqrt{2 \Gamma'}$ as solutions. The resonance center occurs at $X_1\approx -\sqrt{2 \Gamma'}$, the other center at $X_2\approx 0$, and the unstable fixed point at $X_3 \approx \sqrt{2 \Gamma'}$. In this limit, then, the width of the resonance in terms of $\delta \Theta'$ is $\Delta(\delta \Theta') = 4 \sqrt{X_3} \approx 4(2\Gamma')^{1/4}$. 

Since the location of the center of resonance $X_1$ grows in magnitude as $\Gamma'$ increases, the forced resonant eccentricity $\sigma_{res} \propto \sqrt{\Psi'_1} \propto \abs{X_1} \sim \sqrt{2 \Gamma'}$ grows as well. Hence a large $\Gamma'$ limit amounts to a high $\sigma$ limit. How large the eccentricities must be to use this approximation we will quantify momentarily. 

Recall that the width of the resonance is determined by the two crossings of the seperatrix with the $X$ axis (see Figure \ref{fig:contour}). These two crossings in $X$, denoted $X_{\star 1}$ and $X_{\star 2}$, correspond to a minimum and maximum value of $\Psi'$, respectively, for an orbit in resonance for a particular value of $\Gamma'$. These equivalently denote a minimum and maximum value of $\sigma$ for the orbit. We will define the width of the resonance at a initial value of $\sigma$ to be the width of the separatrix assuming $\sigma = \sigma_{\star 1}$ (the minimum value of $\sigma$ for the resonant orbit). This is the same definition of width as we used above for initially circular orbits.

Now, we need to express the width $\Delta (\delta \Theta')$, which is a function of $\Gamma'$, in terms of $\Psi'_{\star 1}$. This involves (approximately) inverting the function 
\begin{align}\label{invert}
\Psi'_{\star 1} & = \frac{X_{\star 1}^2}{2} = \frac{1}{2} (X_3 -2/\sqrt{X_3})^2 \nonumber \\ 
& \approx \Gamma' -2(2 \Gamma')^{1/4} +\frac{\sqrt{2}}{\sqrt{\Gamma'}}
\end{align}
for the relationship $\Gamma' =\Gamma'(\Psi'_{\star 1})$. To obtain the \citet{MustillWyatt} result, we take the limit of large $\Gamma'$ and use $\Gamma' =\Psi'_{\star 1}$. Therefore, the width of the resonance is
\begin{align}
\Delta(\delta \Theta') & \approx 4(2g_m R^2 \sigma^2/Q)^{1/4} 
\end{align}
Following the same analysis as above, we find the critical $m$ for overlap to be
\begin{align}
m& \approx 0.482 (\epsilon_p \sigma)^{-1/5}
\end{align}
leading to the critical period ratio and separation where the first order resonances overlap
\begin{align}\label{nonzeroecc}
\frac{P_2}{P_1} &\lesssim 1+2.08 (\epsilon_p \sigma)^{1/5} 
\end{align}
and
\begin{align}\label{nonzeroecc2}
\frac{a_2-a_1}{a_1}& \lesssim 1.38 (\epsilon_p \sigma)^{1/5}
\end{align}

Note that these results \textit{do not} apply in the case of zero $\sigma$ because of the approximations made (if so, they would incorrectly imply that the widths of the resonances go to zero at zero $\sigma$). When 
\begin{align} 
m \sim  &0.482 (\epsilon_p \sigma)^{-1/5} \gtrsim 0.455 \epsilon_p^{-2/7}
\end{align}
or 
\begin{align}\label{nonzeroecc3}
\sigma \gtrsim 1.33 \epsilon_p^{3/7}
\end{align}
we expect that the overlap criterion in Equation \eqref{nonzeroecc} should approximately hold. This suggests that systems with $\sigma \lesssim (0.013, 0.035,0.093)$ and total planetary mass relative to the host star of $\epsilon_p = 2\times(10^{-5},10^{-4},10^{-3})$, respectively, are adequately described by the resonance overlap criterion for initially circular orbits. 

Equations \eqref{nonzeroecc}, \eqref{nonzeroecc2}, and \eqref{nonzeroecc3} have the same functional form as that derived for the restricted three body problem by \citet{MustillWyatt}, as expected, but with different coefficients. Both the formula derived here and that of \citet{MustillWyatt} are correct to within a factor of a few, as demonstrated by Figure \ref{fig:diagram}.  The curves plotted come from Equation \eqref{nonzeroecc}, and take the form $\log_{10}\{\sigma\} = 5\log_{10}\{P_2/P_1-1\}+\log_{10}\{C^{-5} \epsilon_p^{-1}\} $, where $C=2.08$ (or $2.7$, using \citet{MustillWyatt}). However, the actual slope is closer to 4.6 in the $\epsilon_p = 2\times10^{-5}$ case, and we found it to be closer to 4.2 in the $\epsilon_p = 2\times10^{-4}$ case. Therefore, the scaling of $\frac{P_2}{P_1} -1\propto (\epsilon_p \sigma)^{1/5} $ is only approximate as well. 

There may be a tractable way of obtaining a more accurate first order resonance overlap criterion in the case of nonzero eccentricities. However, such a formula may not be very useful in practice. As we will show in Section \ref{sec:Compare}, the effects of higher order resonances become important even for small eccentricities, and hence the approximation of only considering the first order resonances in deriving an overlap criterion is no longer as valid (though it remains valid at $\sim$ zero eccentricity, where the widths of the higher order resonances are zero). 
%

\section{COMPARISON TO NUMERICAL INTEGRATION} \label{sec:Compare}

Our model of first order mean motion resonances makes clear predictions for the locations of first order mean motion seperatrices and for the location of the chaos resulting from first order resonance overlap. We numerically evolved suites of initial conditions with a range of period ratios, eccentricities, mass ratios, and total mass of the planets, relative to the star using a Wisdom-Holman symplectic integrator \citep{WH1}. We integrate the tangent equations \citep{Licht} concurrently with the equations of motion to determine whether or not the resulting orbits were chaotic. The initial tangent vector used was randomized and normalized to unity. At the end of the integration, the final length of the tangent vector, $d$, is reported. If an orbit is chaotic, $\log{\{ d\}}$ grows exponentially in time, with a characteristic e-folding time equal to the Lyapunov time. If the orbit is 
regular, we expect polynomial growth in time: $\log{\{ d\}} = p \log{\{ t_{final}\}}$, where $p$ is the polynomial exponent.  The estimated Lyapunov time $T_{Ly}$, then, is $ T_{Ly} = t_{final}/\log{\{ d\}}$, where $t_{final}$ is the length of the integration. 

After an ensemble of integrations, we expect an approximately bimodal distribution of Lyapunov times.  The estimated Lyapunov times of \emph{regular} orbits should be sharply peaked at approximately $10-100$ times shorter than the integration time (assuming $p \lesssim 5$). The estimated Lyapunov time of \emph{chaotic} orbits will fall to shorter times than this peak. Long enough integrations clearly differentiate between the two, though we cannot prove that an orbit is regular through numerical integration alone. Our integrations were $10^6$ days long unless noted otherwise. This is $\sim 3,000$ orbits of the outer planet, the orbital period of which was fixed at one year. We tested how the integration time could affect our results by comparing the output from integrations of $10^6$, $10^7$ and $10^8$ days and found that $10^6$ days provided reliable results for the orbits studied; see the Appendix for more details. 

A symplectic corrector \citep{Touma} identical to the one used in Mercury \citep{Chambers} was implemented to improve the accuracy of the integrations. With $\epsilon_p \sim 10^{-5}$,  the typical fractional energy error was smaller than $10^{-8}$ for orbits that do not suffer close encounters. We may not be resolving close encounters with high fidelity, but, orbits that experience close encounters and strong gravitational scatterings are certainly chaotic, and hence for our purposes it is less important to resolve the close encounters accurately than to identify that the orbits are chaotic (by measuring a short Lyapunov time).  

Symplectic mappings are known to introduce chaotic behavior when commensurabilities occur between the timestep frequency $2\pi/\Delta t$ and orbital frequencies of the physical problem \citep{WH2}. In practice, this chaotic behavior is confined to an exponentially small region of phase space as long as the shortest timescale is resolved by a factor of $\sim 20$ \citep{RH}. Our time step of $\sim1$ day, for orbits with a minimum period of half of a year, ensures that we do not need to worry about integrator chaos.  
\subsection{Resonance Widths as a Function of Orbital Elements}
\begin{figure}
\begin{center}
\caption{Chaotic structure of phase space as a function of period ratio of the planets and the weighted eccentricity $\sigma$. In these plots, $e_2(0)=0$, so $\sigma=e_1$. Darker colors indicate shorter Lyapunov times. The three panels show the results for different initial values of the mean anomaly of the inner planet, $M_1$. The color scale is cut off at a value of $T_{Ly} = 3P_1$, and orbits with shorter Lyapunov times than this are in black. Any orbits with Lyapunov times longer than the upper value of the color scale are in yellow. The seperatrices for each resonance as predicted by the analytic theory are shown in grey.  }\label{fig:overlap1}
\includegraphics[width=3.5in]{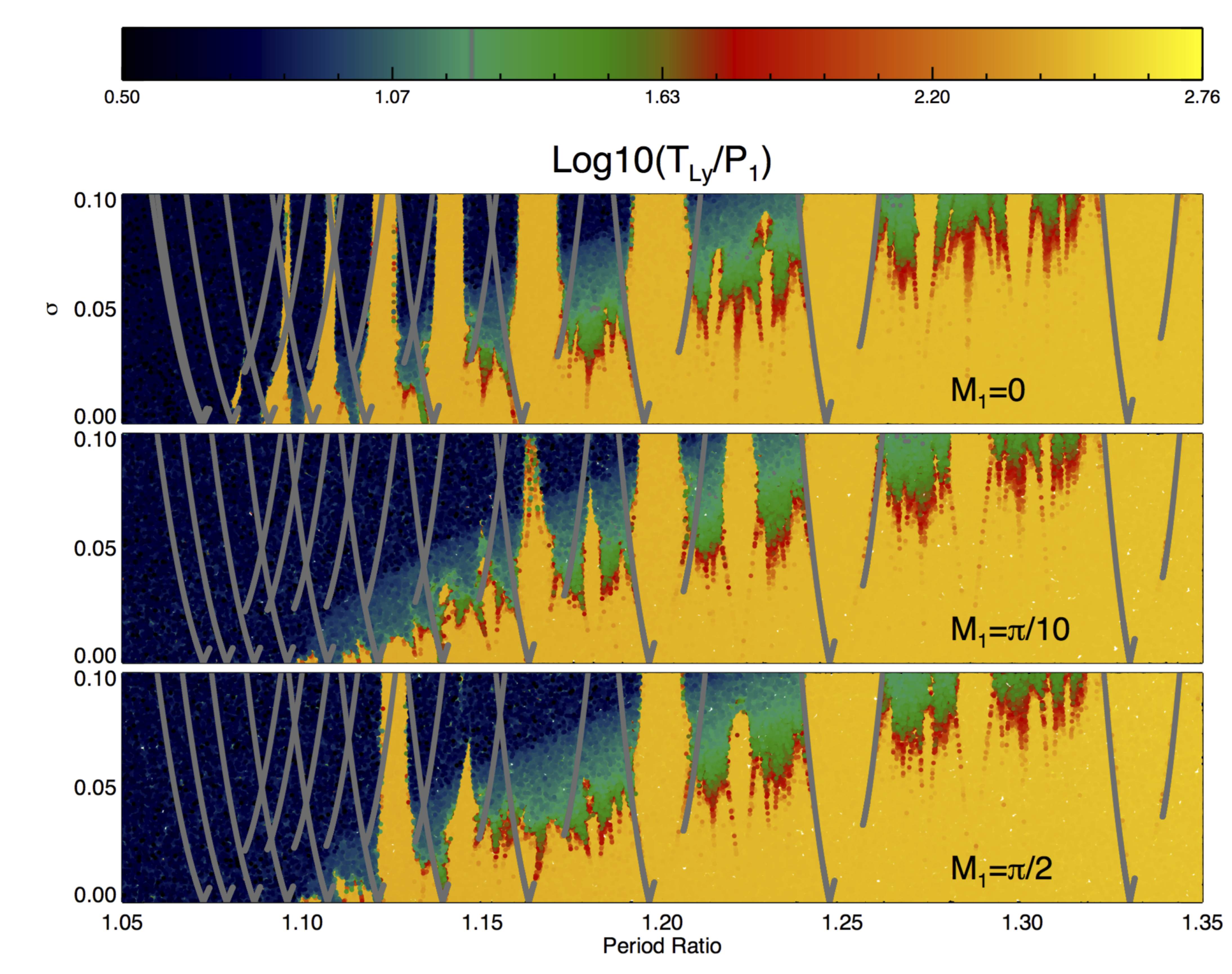}

\end{center}
\end{figure}

The analytic model for first order resonances can be used to predict widths as a function of $\sigma$ and period ratio as in Figure \ref{fig:MRtwo}, or, for a fixed $\sigma$, the widths as a function of period ratio and initial orbital angles as in Figure \ref{fig:ChainSep}. In Figure \ref{fig:overlap1} we show the results of our numerical integrations as a function of $\sigma$ and period ratio for three different values of the initial mean anomaly of the inner planet, $M_1(0)$. In particular, the masses of the planets are held fixed at $\epsilon_1 = \epsilon_2 = 10^{-5}$, $e_2(0) = 0$, and all other orbital angles are fixed at zero. The coloring indicates the measured Lyapunov time of the orbits, in units of the initial inner planet's period; darker colors reflect shorter times. In each case, we have overplotted the predicted resonance widths based on the analytic model.

The upper panel shows the case of $M_1(0) = 0$. A regular (yellow) region appears for every first order resonance until the first order resonances are completely overlapped, at period ratios near unity, where a large chaotic zone appears that extends to zero eccentricity. However, it is clear that higher order resonances are important in explaining much of the chaotic behavior we see at higher eccentricities. For example, even though the 5:4 at $P_2/P_1 = 1.25$ ($m=4$) and the 4:3 at $P_2/P_1 = 1.\bar{3}$ ($m=3$) first order resonances are far from overlapping, one can see both the chaotic separatrices of and regular regions associated with the two third order resonances between them (though the second order resonance between them appears to be chaotic). As the period ratio nears unity, overlap between second order and first order resonances starts to occur, and hence the region around the first order seperatrices becomes chaotic even though the first order resonances themselves do not overlap with each other. It is clear that the chaos is not confined to a region where the first order resonances overlap; this is why we believe that a first order resonance overlap criterion as a function of eccentricity may not be very applicable in practice. 

\begin{figure} 
		\begin{center}
		\caption{Map of the chaotic structure as a function of $M_1$ and period ratio, with all else held fixed. Yellow regions embedded in the chaotic web corresponds to resonant islands. The islands in a particular chain appear disjoint because there is an unstable fixed point in between each one. Chaos appears first near the unstable fixed points. Note also that all other orbital angles are initially zero, so the initial value of $\phi(0)$ depends only on $M_1$. }\label{fig:Illustrate}
	\includegraphics[width=3.5in]{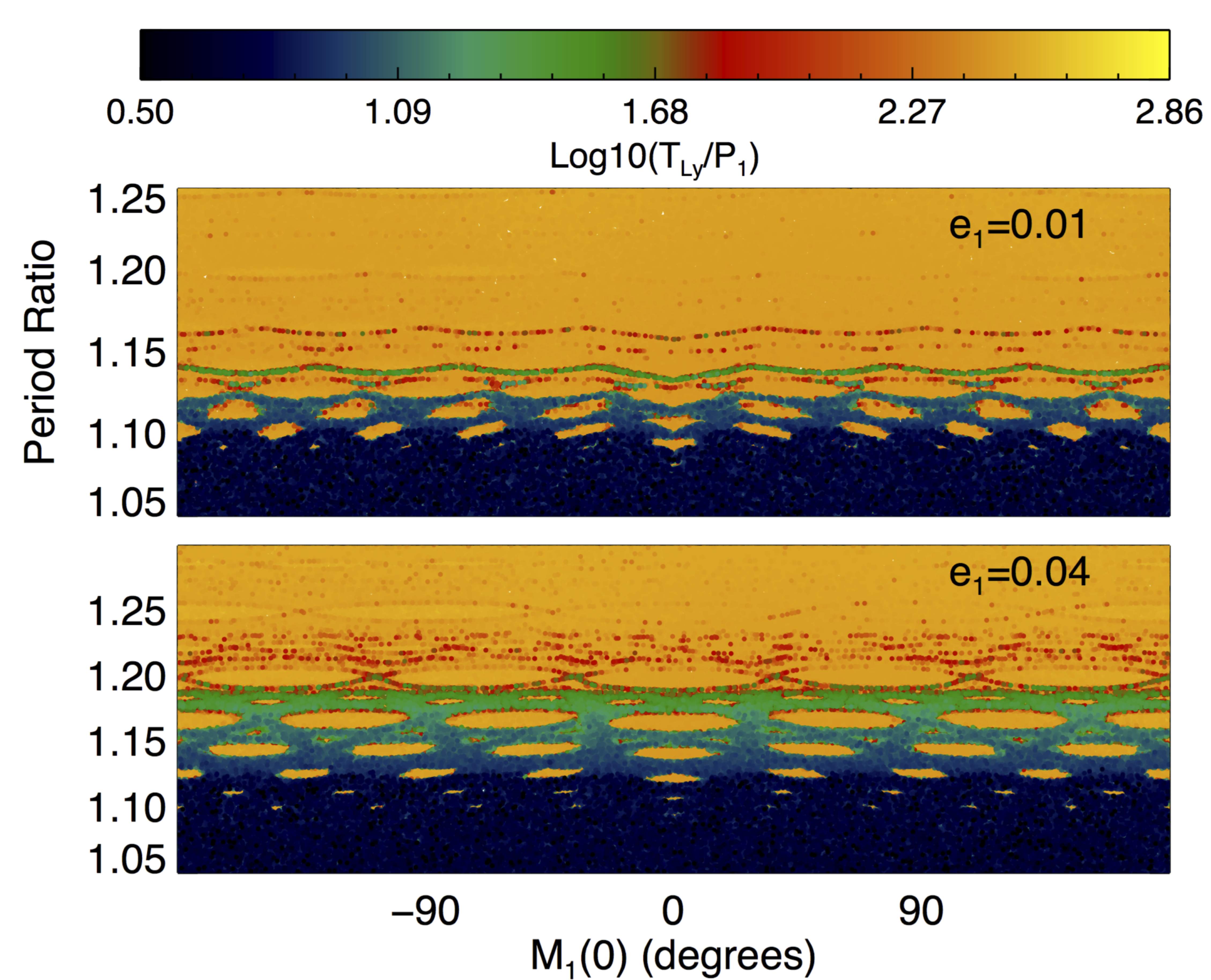} 

	\end{center}
\end{figure}

In the lower two panels, the same picture emerges, but not every first order resonance appears as a yellow region - in some cases, the first order resonances are entirely chaotic when they were entirely regular at $M_1(0)=0$. The reason for this is demonstrated in Figure \ref{fig:Illustrate}, which shows  the chaotic regions of the phase space as a function of period ratio and $M_1$, with all other initial orbital elements held fixed. The plots in the figure, which correspond to nonzero eccentricities, agree with the predictions of the model as to the number of and location of resonance islands for a particular value of $m$ (and should be compared with Figure \ref{fig:ChainSep}). 

Recall that the resonant overlap criterion only predicts the extent of the chaotic zone where $\sim$\textit{all} orbits will be chaotic. The chaotic zone at larger period ratios contains many regular islands; as a result, we refer to it as the chaotic web. These regular islands correspond to first order mean motion resonances that are only weakly overlapped with neighboring first (or second) order resonances. The resonant islands at a particular period ratio are smaller in the lower panel (where $e_1$ is larger) compared to the upper panel. This is because the resonance widths grow with eccentricity and consequently the overlap becomes more extreme.

It is only in the case of $M_1=0$ that islands associated with every $m$ get sampled (think of taking a vertical slice in Figure \ref{fig:Illustrate} at $M_1=0$).  Other values of $M_1$ do not pass through as many resonant islands.  Using Equation \eqref{formula}, with $e_2(0)=0$ and all angles zero except $\lambda_1$ yields the location of the resonant islands in terms of $\lambda_1$:
\begin{align}\label{location}
\lambda_1 \approx M_1 &= -\frac{2k\pi}{m},
\end{align}
with $k$ an integer ranging from $0$ to $m-1$. We see that the case $M_1=0$ corresponds to the center of a resonant island for every $m$ ($k=0$). The center of the islands for the $6:7$ resonance at $P_2/P_1 \sim 1.1667$ ($m=6$), for example, appear at $M_1 = -\frac{k}{3}\pi = (0,\pi/3,2\pi/3,\pi,4\pi/3,5\pi/3)$. A value of $M_1 = \pi/2$ falls directly in between two of these islands, right at the unstable fixed point. It makes sense, then, that the $6:7$ resonance appears entirely chaotic when $M_1 = \pi/2$, while it appears regular for $M_1=0$.  The dependence of the location of the resonant islands on $m$ and $M_1$ explains many of the features in Figure \ref{fig:overlap1}.  It is clear that the case of $M_1=0$ (and all other angles equal to zero) is a special case; even the small change to $M_1(0) = \pi/10$ leads to significant changes in the structure of the chaotic zone.

Another interesting feature of Figure \ref{fig:Illustrate} is that the locations of the resonances at $M_1(0)=0$ are consistently shifted to smaller period ratios compared to neighbors in the same resonant chain. However, the analytic model developed in Section \ref{sec:HamDevelop} predicts no such behavior. This effect is caused by the transformation between real coordinates (those used in the numerical integration) and averaged variables (those used in the analytic development). The averaging step was necessary to remove the short period terms in the disturbing function. In the Appendix, we derive the canonical transformation between the two sets of variables and show that in the case of small eccentricities, the shift is in the direction observed and it is maximized when $M_1=0$. In Figure \ref{fig:overlap1} we have corrected the theoretical curves in the case of $M_1=0$ to account for this transformation (but only using the zero eccentricity terms). There is still a slight deviation between the analytic curves and the numerically determined seperatrices at period ratios near unity and nearly zero eccentricities. This discrepancy could presumably be removed if we carried the transformation to second order in the masses.

In summary, the initial value of the angles do not greatly affect the overall extent of the chaotic web in terms of period ratio (as shown in Figure \ref{fig:Illustrate}). The locations of the resonant islands in this web do depend on the orbital angles, but in a way that is predicted by the analytic theory and confirmed numerically. Although $e_2$ and $\Delta \varpi$ were initially zero for all of these integrations, the individual values of $e_1$, $e_2$, and $\Delta \varpi$ do not matter. Only the weighted eccentricity $\sigma$ (and the generalized longitude of pericenter $\psi$) matter. We confirmed that the predicted boundaries of the resonance in terms of $\sigma$ matched well with those observed numerically in the case of $e_1$ variable, $e_2=0.02$, and $\Delta \varpi = \pi/2$: see the Appendix for details.

\subsection{Planetary Mass Ratio Dependence}\label{sec:MRcompare}
\begin{figure}
	\begin{center}
	\includegraphics[width=3.5in]{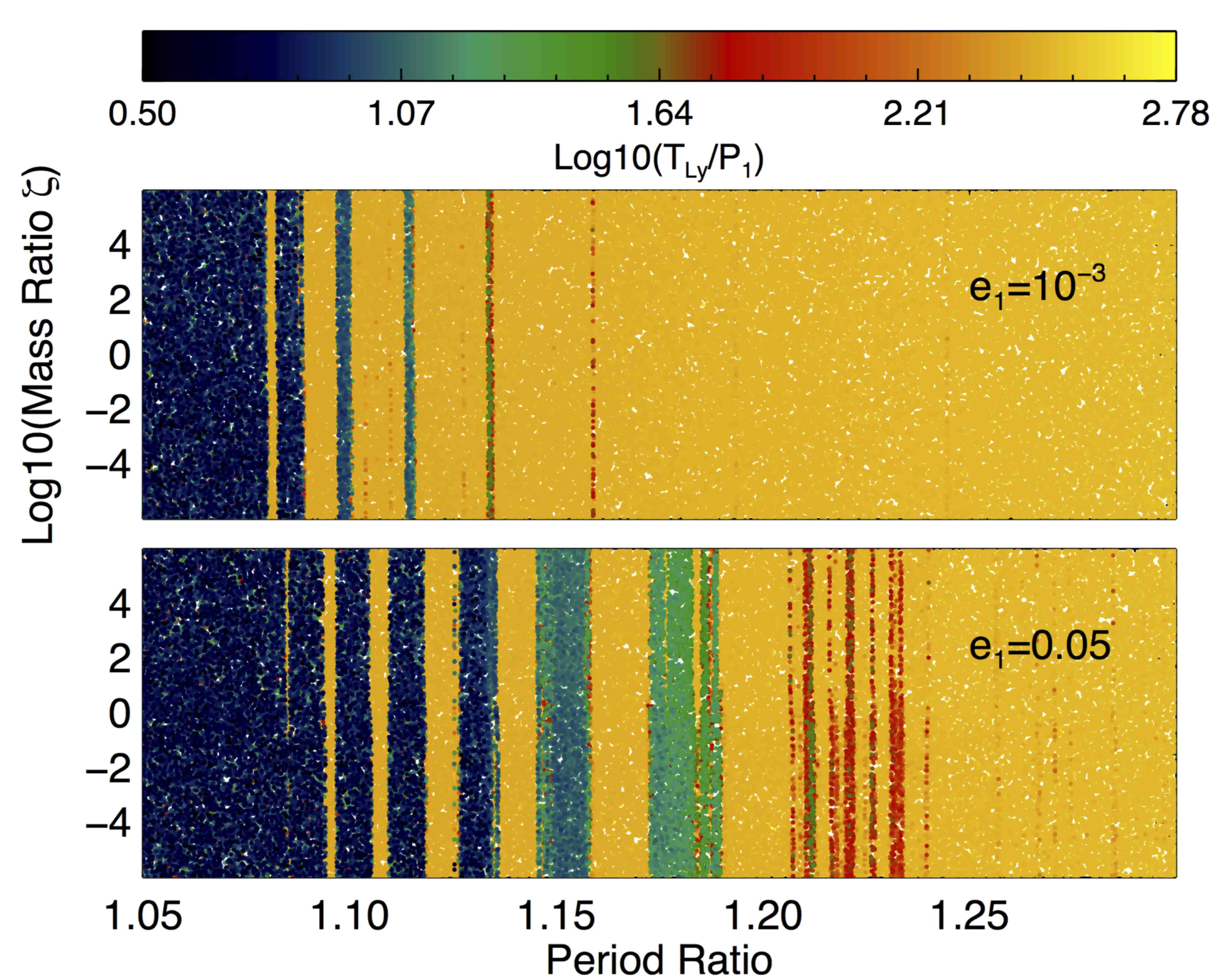}
	\caption{Map of the chaotic regions of phase space as a function of period ratio and mass ratio of the planets. The color scale reflects the numerically estimated Lyapunov time, relative to the initial period of the inner planet. The effect of the mass ratio of the planets on the chaotic zones is negligible, even at higher eccentricities. These integrations used initially circular planets with $\epsilon_p= 10^{-5}$. Only the initial period ratio and mass ratio of the planets is varied. All orbital angles are set to zero (the initial orbital angles chosen will affect the location and widths of these bands of chaos).}
	\label{fig:widthNumer}
	\end{center}
	\end{figure}

One of the main results of the analytic work is that the widths of first order mean motion resonances should be independent of the mass ratio of the planets, for a fixed $\epsilon_p$, especially for resonances with higher values of $m$. This implies that the width of the chaotic zone due to overlap of first order mean motion resonances is independent of $\zeta$.  We confirm numerically that the structure of the chaotic zone does not change significantly over 12 orders of magnitude in $\zeta$; the results are shown in Figure \ref{fig:widthNumer} for both $\sim$zero eccentricity ($e_1 = 10^{-3}, e_2=0$) and nonzero eccentricity ($e_1 = 0.05, e_2=0$). These plots can be thought of as horizontal slices at a particular $\sigma$ in Figure \ref{fig:overlap1}.

At very small eccentricities, the chaotic zone is due almost entirely to first order mean motion resonance overlap, and so it makes sense that we see no significant dependence on $\zeta$.  However, it is surprising that the mass ratio is relatively unimportant even at higher eccentricities and at larger period ratios, where the chaos we are observing is due to second and third order mean motion resonance overlap. These results suggest that the widths of higher order resonances, in the case of two massive planets on close orbits, can be approximated analytically using the restricted three body problem. We defer an investigation of this to future work.
 
\begin{figure} 
		\begin{center}
		\caption{Chaotic structure of phase space as a function of period ratio of the planets and the weighted eccentricity $\sigma$. In these plots, $e_2(0)=0$, so $\sigma=e_1$. We set $\epsilon_p=2\times10^{-5}$ and $\zeta=10$, with all angles set to zero initially except $M_1=\pi/2$. This plot is completely analogous to and should be compared to the bottom panel of Figure \ref{fig:overlap1}; the only difference is the value of $\zeta$ used. Changing the planetary mass ratio makes no significant changes to the structure of the chaotic region.
		}\label{fig:overlapMasses}
	\includegraphics[width=3.5in]{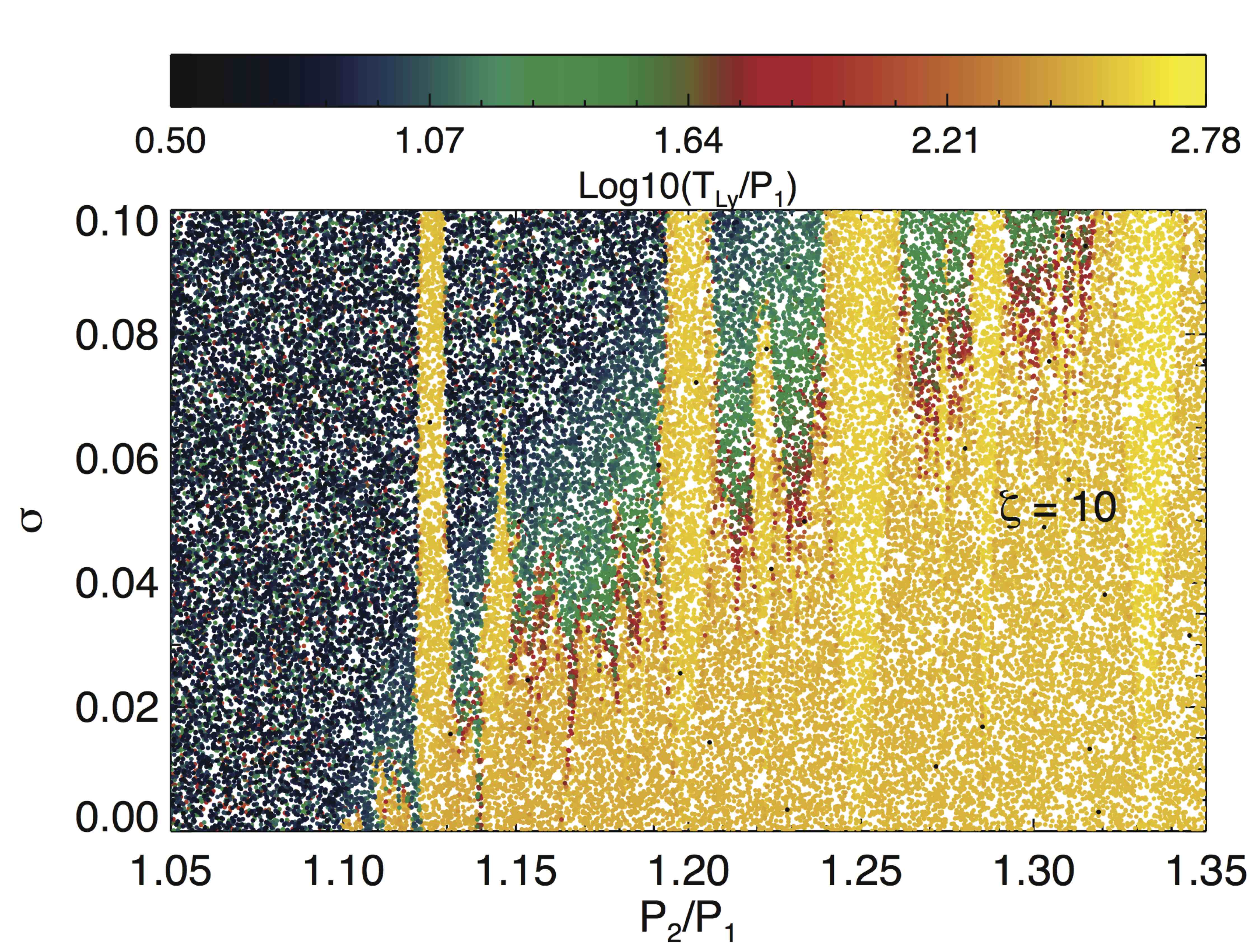} 

	\end{center}
\end{figure}

 The independence of the structure of the chaotic region on the mass ratio is even more striking when we consider a wide range of eccentricities.  Figure \ref{fig:overlapMasses} shows the chaotic regions of phase space for the same initial conditions used to create the bottom panel in Figure \ref{fig:overlap1}. The only parameter that has changed is the mass ratio of the planets $\zeta$, which increased from 1 to 10. The total mass of the planets is unchanged. Figure \ref{fig:overlapMasses} is almost indistinguishable from the bottom panel of Figure \ref{fig:overlap1}. Tests with $\zeta=100$ and $\zeta=0.1$ showed similar behavior. It is clear that the mass ratio of the planets matters very little in determining the structure of the chaotic zone in this regime ($P_2/P_1 \lesssim 1.5- 2, \sigma \lesssim 0.1$).


\subsection{Effect of the Total Mass of the Planets Relative to the Mass of the Host Star}
\begin{figure}
	\begin{center}
	\includegraphics[width=3.5in]{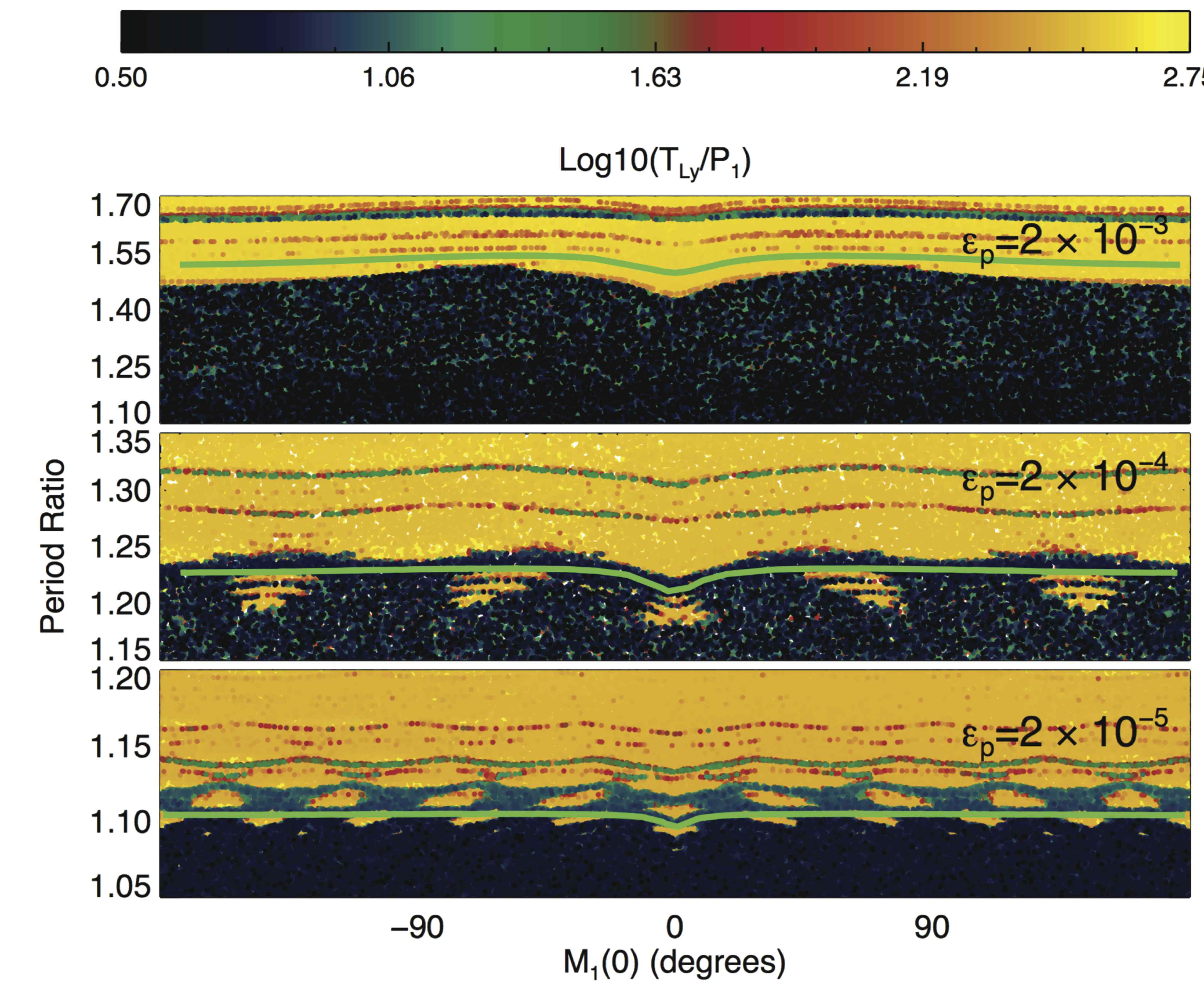}
	\caption{The effect of the total mass of the planets relative to the mass of the host star, $\epsilon_p$, on the chaotic structure of phase space for close orbits. \textit{Top:} $\epsilon_p = 2\times 10^{-3}$, \textit{Middle:} $\epsilon_p = 2\times 10^{-4}$, \textit{Bottom:} $\epsilon_p = 2\times 10^{-5}$. For more massive planets, relative to the star, the first order resonances are more overlapped, and less of the phase space at low eccentricity and period ratios near unity is regular.  In each plot, the thick neon green line corresponds approximately to the Hill boundary. For low mass planets there are more resonant islands.}
	\label{fig:totalmass}
	\end{center}
	\end{figure}

The resonance widths grow with the total mass of the planets, relative to the host star, as $\epsilon_p^{2/3}$, and so we expect that the overall extent of the chaotic web should grow with $\epsilon_p$ as well. Figure \ref{fig:totalmass} shows the chaotic structure of the phase space in the period ratio - $M_1$ plane for three different values of $\epsilon_p$ (the bottom panel, with $\epsilon_p = 2\times10^{-5}$) is the same as upper panel of Figure \ref{fig:Illustrate}). The eccentricity of the inner orbit $e_1$ was fixed at 0.01, $e_2$ was fixed at 0. The discussion accompanying Figure \ref{fig:Illustrate} explained why resonant islands appear at all, and how their sizes depend on the eccentricities. Here we see how they change as the total mass of the planets grows. 

At the highest masses shown ($\epsilon_p=2\times10^{-3}$), there are no resonant islands at all, since the widths of the resonances are large enough to completely overlap. Presumably the abrupt transition at period ratios of $\sim 1.45$ to a mostly regular phase space happens because the 3:2 resonance does not overlap much with the second order resonance above it (the 5:3) resonance, while it does with the second order resonance below it (the 7:5) since the 7:5 is closer.  The planets would have to be more massive than Jupiter or more eccentric in order to merge the chaotic seperatrices of the 3:2 and the 5:3 resonance. 

The effect of the transformation between real and averaged coordinates grows $\propto \epsilon_p$ (see Equation \eqref{shiftFormula}). As a result, the shift at $M_1=0$ is much more extreme in the $\epsilon_p=2\times10^{-4}$ case compared to the $\epsilon_p=2\times10^{-5}$ case.

Finally we point out that for low mass planets on low eccentricity orbits, the Hill boundary (shown in green) lies within the chaotic web. The implications of this will be discussed in Section \ref{sec:Discuss}.

\subsection{Resonance Overlap Scaling}
\begin{figure*}
		\begin{center}
		\caption{ Comparison of the size of the chaotic zone for initially circular orbits as predicted by the resonance overlap criterion (Equation \eqref{crit_overlap}, green line) to that determined by numerical integrations. The black points represent the location of chaotic orbits in terms of the initial value of $a_1$ and the total mass of the planets relative to the mass of the host star, $\epsilon_p$.  In red is the Hill boundary (Equation \eqref{Hill2}); the resonance overlap criterion is a stricter criterion for low mass planets. Only $a_1$ and $\epsilon_p$ were varied within each panel, with $M_1$ varied between panels. All other angles were fixed at zero.  We have not applied the correction between numerical coordinates and averaged coordinates in this figure. The correction is small and will mainly affect the results when $M_1=0$. 
 }
	\includegraphics[width=6.5in]{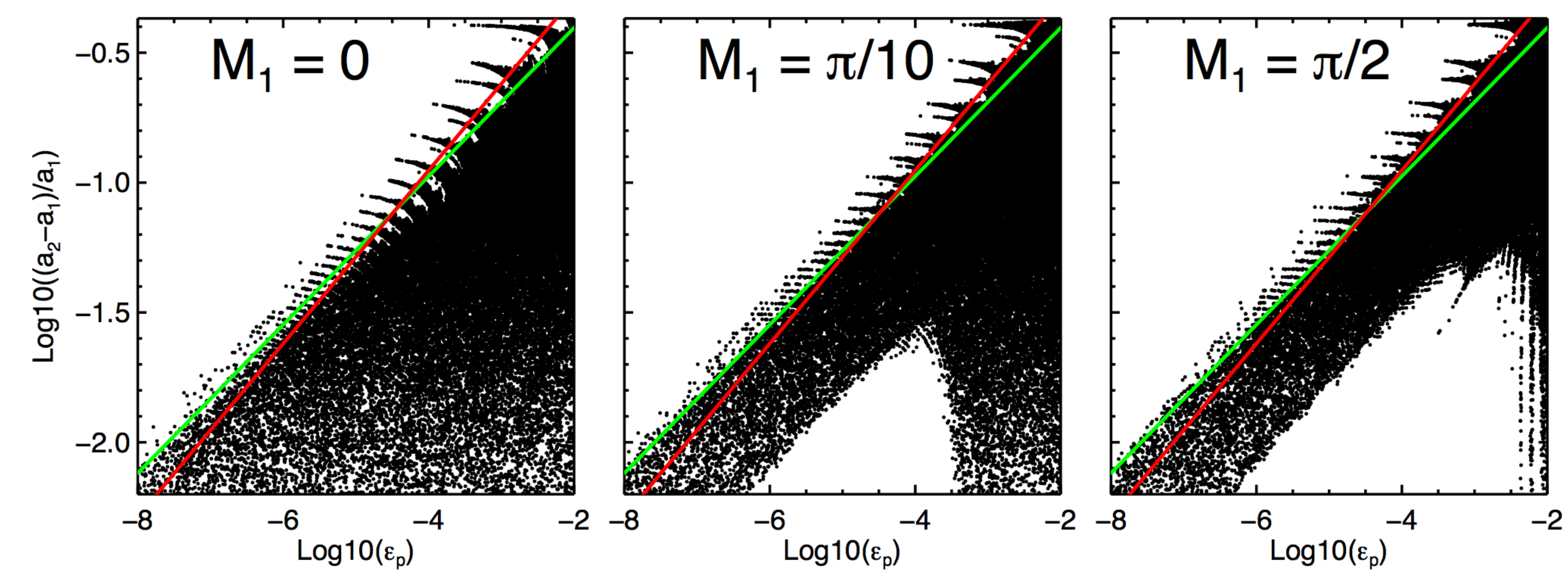} 
\label{fig:IllustrateScaling}
	\end{center}
\end{figure*}

How well do the derived resonance overlap criteria given in Equation \eqref{crit_overlap} and Equation \eqref{nonzeroecc2} predict the size of chaotic zone in practice?  In Figure \ref{fig:IllustrateScaling} we show, in terms of the initial value of $\log_{10}{\{a_2/a_1-1\}}$ and $\log_{10}{\{\epsilon_p\}}$, the location of initially circular orbits we determined to be chaotic numerically. Only the total mass of the planets relative to the mass of the host star $\epsilon_p$ and the semimajor axis of the inner planet were varied. Recall that the criterion predicts, given $\epsilon_p$, the extent of the chaotic zone where there are no regular orbits (aside from those protected by the 1:1).  The derived formula for the boundary of the chaotic zone closely tracks the numerically determined edge across a wide range of $\epsilon_p$, regardless of the initial values of the orbital angles. When the mass of the planets is larger, the simple scaling appears to underestimate slightly the extent of first order resonance overlap for initially zero eccentricity orbits. A slight discrepancy was expected here due to higher order terms, as discussed in Section \ref{sec:Resoverlap}. Interestingly, the Hill boundary appears to track the extent of the chaotic zone for more massive planets, at least when $M_1(0)$ is further from zero. 

For intermediate mass planets, with initial orbital configurations $M_1 \neq M_2$,  a regular region appears corresponding to the 1:1 mean motion resonance. The 1:1 region will be explored in future work (Payne et al, in prep).

	\begin{figure}
		\begin{center}
		\caption{ Comparison of the size of the chaotic zone for initially eccentric orbits as predicted by the eccentric resonance overlap criterion (Equation \eqref{nonzeroecc2},  purple line) to that determined by the analysis of \citet{MustillWyatt} (the green line) and the results of numerical integrations. The estimate of the size of the chaotic zone predicted by the initially circular resonance overlap criterion (Equation \eqref{crit_overlap}, cyan line) is also shown. It is clear that first order resonance overlap alone cannot account for all of the chaos shown, particularily at lower masses. Within each panel, only the initial orbital separation of the planets, $e_1$, and $M_1$ vary. $M_1$ was chosen randomly.  All other angles were fixed at zero, and $e_2=0$. The different panels correspond to different values for $\epsilon_p$, $\zeta$ was fixed at 1.
 }
	\includegraphics[width=3.5in]{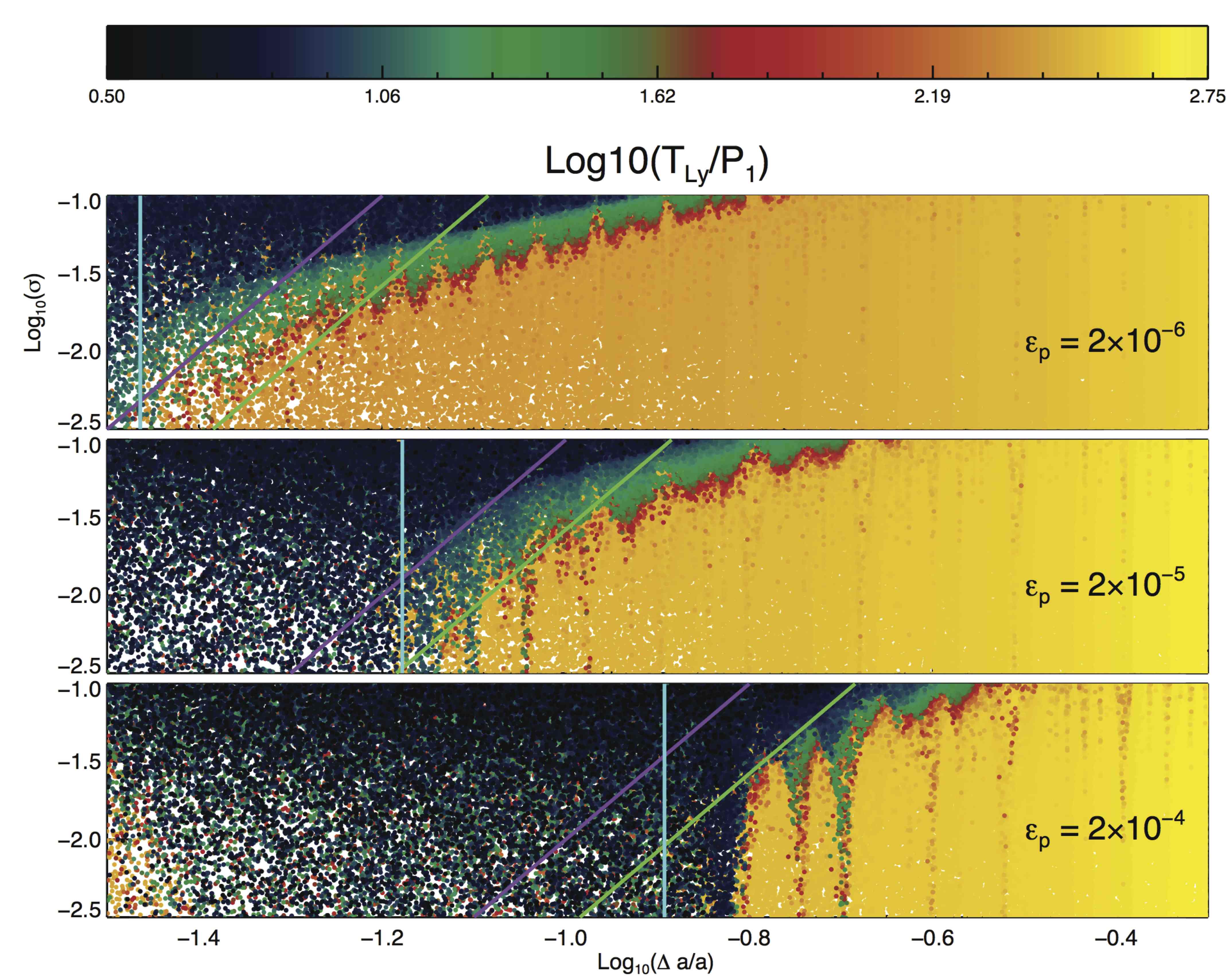} 
\label{fig:MW}
	\end{center}
\end{figure}

Figure \ref{fig:MW} shows the boundary of the chaotic zone as predicted by the overlap criterion for initially eccentric orbits  (Equation \eqref{nonzeroecc2}) on the $\log_{10}\{\sigma\}-\log_{10}{\{a_2/a_1-1\}}$ plane for three different values of $\epsilon_p$. We also show the estimate of \citet{MustillWyatt} (which lies parallel to the estimate derived in this work, both have slopes of 5 but differ by a constant coefficient). These lines apply in the large $\sigma$ regime, after they cross the overlap boundary predicted for initially circular orbits (the vertical line). The \citet{MustillWyatt} coefficient does fit better than the one derived in Section \ref{sec:resEcc}. However, it is clear that the first order resonance scaling as a function of $\sigma$ is approximate. It tracks the boundary of the chaotic zone only for a small range of $\sigma$. Once $\sigma$ is large enough, second and third order resonances become important, and the slope of the edge of the chaotic zone on the $\log_{10}\{\sigma\}-\log_{10}{\{a_2/a_1-1\}}$ plane changes significantly from the estimated value of 5. 
\section{DISCUSSION}\label{sec:Discuss}

\subsection{Chaotic Dynamics and the Lifetime of Chaotic Orbits}

In order to use the resonance overlap criterion as a stability criterion, we need to ensure that the orbits in the chaotic zone are indeed Lagrange unstable. \citet{Rein} found that in a region of phase space similar to that  inhabited by the Kepler 36 system (period ratio close to 6:7, $e_1=0,e_2<0.04$) orbits that were chaotic also eventually experienced at least a $10\%$ variation in semimajor axis (compared to the initial value) and hence were Lagrange unstable. A smaller subset of these initial conditions led to ejections of the outer planet. Hence we might expect that orbits in the chaotic zone are indeed Lagrange unstable on short timescales. Gladman also found that the majority of the chaotic orbits were Lagrange unstable, but that near the edge of the chaotic web orbits appeared to be relatively quiescent for many thousands of Lyapunov times, despite having very short Lyapunov times (on the order of the synodic period of the planets) \citep{Gladman}.

		\begin{figure}
	\begin{center}
	\includegraphics[width=3.5in]{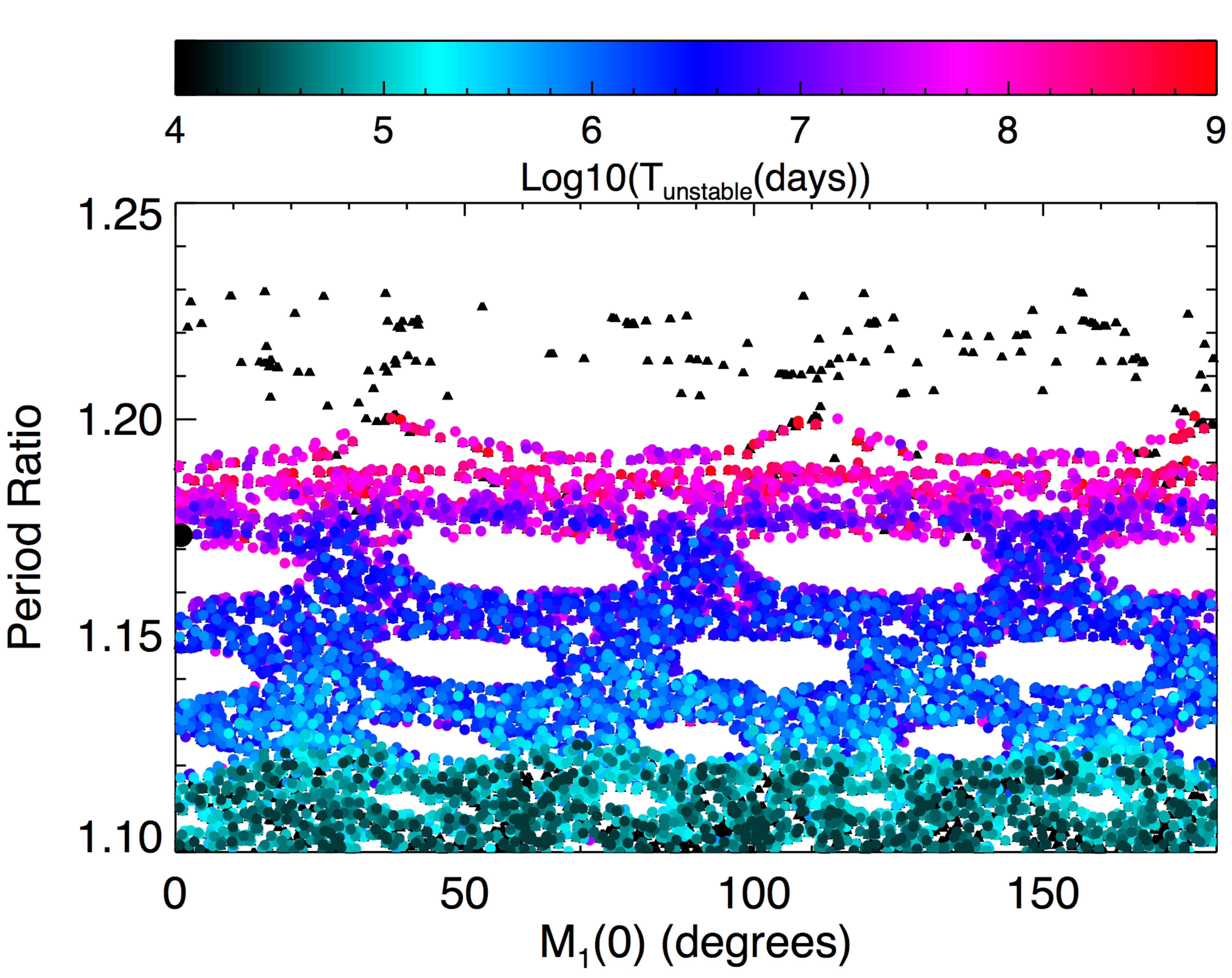}
	\caption{The location of chaotic orbits (in black triangles, as determined by integrations of $10^6$ days) compared to those that show Lagrange instabilities within $10^9$ days (in colored circles). The initial conditions used here are the same as those used to create the lower panel of Figure \ref{fig:Illustrate}. The location of unstable orbits closely track the chaotic region of phase space - underneath nearly every colored point is a black triangle. Note that regions with no points plotted were integrated and found to be both regular and long-lived.  For reference, period ratios larger than $\sim 1.148$ are Hill stable, so the chaotic web extends past the Hill boundary. The large black dot corresponds to the initial condition shown in Figure \ref{fig:evol} ($M_1(0)\sim 0, P_2/P_1\sim1.17$). }
	\label{fig:unstable2}
	\end{center}
	\end{figure}

To determine the stability of the orbits in the chaotic zone,  we studied the evolution of a subset of our initial conditions. The majority of the initial conditions were integrated for $10^8$ days. Chaotic orbits which did not show instability on this timescale were integrated for $10^9$ days. Energy was conserved in these integrations to within a part in $10^{10}$ for orbits not resulting in strong scattering events. Orbits where the semimajor axes changed by more than 5\% from their initial values were classified as Lagrange unstable; the first instance that $\abs{a(t)-a(0)}/a(0)>0.05$ is satisfied is the instability time. We find that many of the chaotic orbits are indeed Lagrange unstable on short timescales, regardless of whether they satisfy the Hill criterion.  Figure \ref{fig:unstable2} shows the results of a test using the same initial conditions as the bottom panel of Figure \ref{fig:Illustrate}. The location of the unstable orbits (in colored points) closely track the structure of the chaotic web (in black triangles). A comparison of Figure \ref{fig:unstable2} and Figure \ref{fig:Illustrate} illustrates how similar the two are; underneath almost every colored circle is a black triangle. Note that the resonance overlap criterion does not immediately apply to this case, as the eccentricities are nonzero. 

All unstable orbits are chaotic, but it is also clear that a small fraction of the chaotic orbits do not show instabilities during the $10^9$ day integrations. As expected based on Gladman's work, these orbits are grouped near the edge of the chaotic web; all satisfy the Hill criterion. Since the timescale to exhibit instability is consistently longer nearer the edge of the chaotic web, we predict that chaotic orbits that have not shown instabilities in $10^9$ days will reveal erratic motion in longer integrations (though if the instability timescale is longer than $\sim 10^{12}$ days the system is effectively long-lived).  The dependence of an instability timescale on the initial location of an orbit in the chaotic zone has been observed  \citep{MH2,Kepler36} and interpreted as a dependence of the chaotic diffusion coefficient on the orbital elements. Note that chaotic orbits near the edge of the chaotic web (in period ratio) are those that satisfy the Hill criterion by the largest amount, a quality that has been found to correlate with the timescale to exhibit Lagrange instabilities \citep{Kepler36}. All of the initial conditions leading to ejections or extremely wide two planet systems within $10^9$ days fail the Hill criterion.
		\begin{figure}
	\begin{center}
	\includegraphics[width=3.5in]{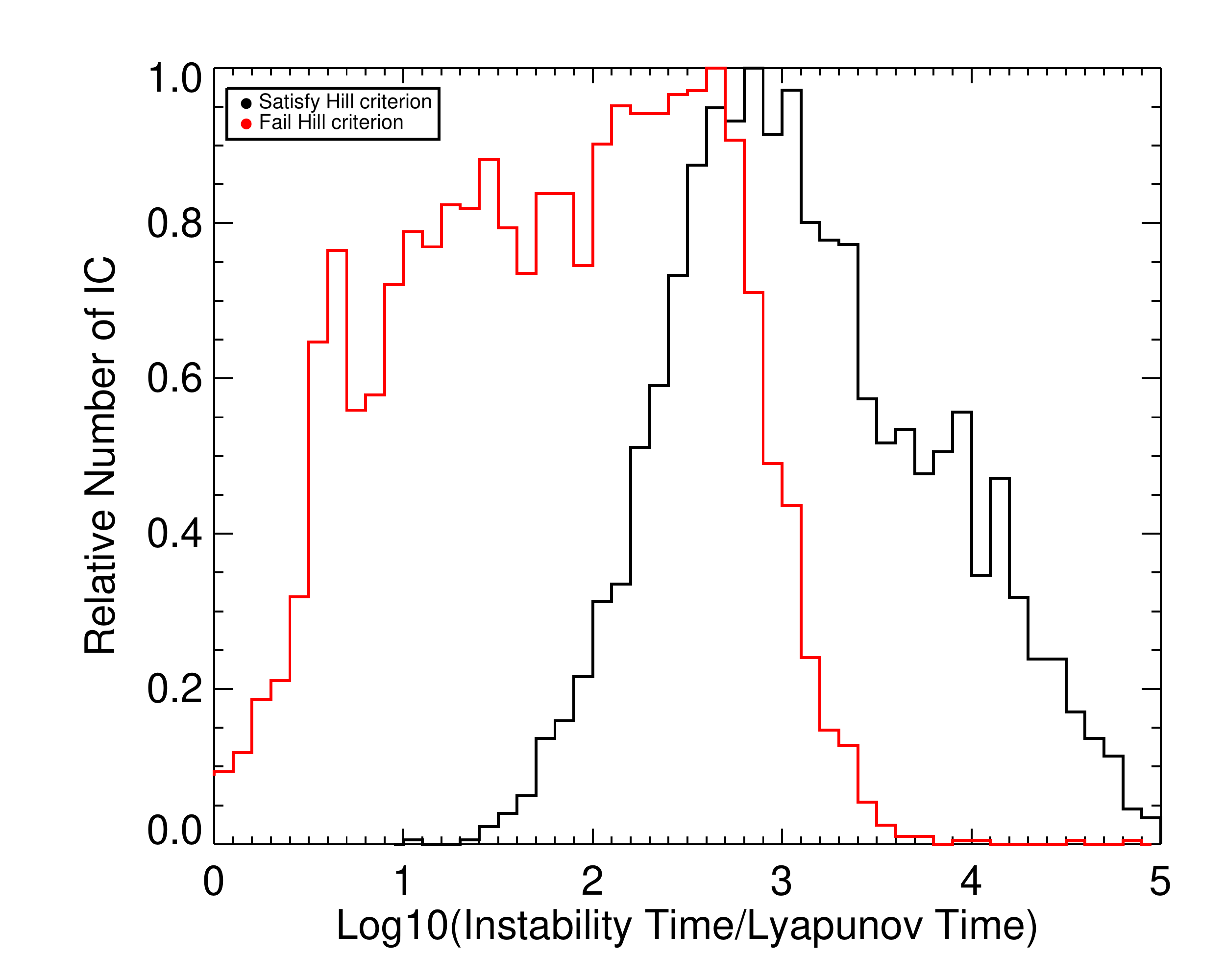}
	\caption{The distribution of the ratio of the Lagrange instability time to the Lyapunov time of the unstable orbits. Orbits which fail the Hill criterion exhibit instabilities in significantly fewer Lyapunov times compared to those which satisfy the criterion. There is not a clear relationship between Lyapunov time and instability time. }
	\label{fig:histo22}
	\end{center}
	\end{figure}
	
We find that the estimated Lyapunov time of these orbits are $0.3-30$ times the initial synodic period of the pair, similar to Gladman's result. Note that both the estimated Lyapunov time and the synodic period of Lagrange unstable orbits change can change as orbital instabilities set in. The distribution of the ratio of the instability time to the Lyapunov time is shown in Figure \ref{fig:histo22} for the unstable initial conditions, showing that indeed the Lyapunov time is orders of magnitude shorter than the instability time.

How do trajectories evolve within this chaotic web once instabilities set in? Although the chaotic zones at different period ratios look disjoint in the plots of period ratio vs. $\sigma$ (for example, Figure \ref{fig:overlap1}), other slices of the phase space (particularly in period ratio vs $M_1$, e.g. Figure \ref{fig:Illustrate}) suggest that these zones are part of a larger connected chaotic region. The similar Lyapunov times across the chaotic regions also hints that the same mechanism is responsible for the chaos. In either case, however, we are looking at a projection of the chaotic zone onto a two dimensional subspace and so we cannot claim based on these figures that the chaotic web is indeed connected, making up a single chaotic zone.

\begin{figure}
	\begin{center}
	\includegraphics[width=3.5in]{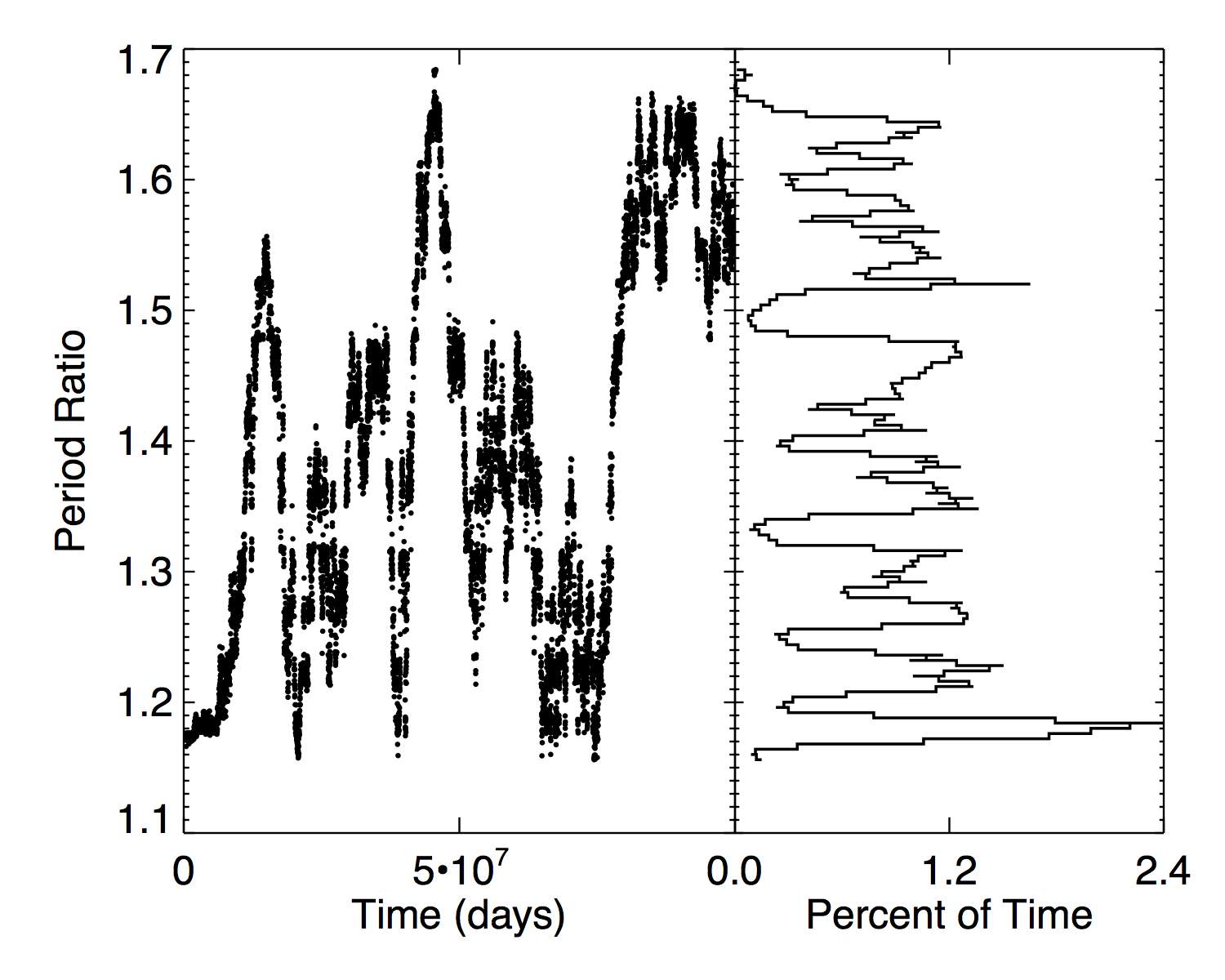}
	\caption{Evolution of the period ratio from a single integration of two planets in a Hill stable, but Lagrange unstable, orbital configuration. The distribution of the period ratio is what one expects from resonance overlap of the first and second order mean motion resonances.   The location of this initial condition in the chaotic web is shown in Figure \ref{fig:unstable2} and Figure \ref{fig:dL}. The maximum fractional deviation in the energy for this integration is max$\abs{(E(t)-E(0))/E(0)}\sim3\times10^{-10}$, and the estimated Lyapunov time is 21,144 days, or $\sim 67 P_1(0)$. }
	\label{fig:evol}
	\end{center}
	\end{figure}

 Figure \ref{fig:evol} shows the time evolution of the period ratio of a single Hill stable initial condition. The trajectory indeed wanders throughout the chaotic web, which extends over a large range of period ratios. The distribution of the period ratio is what one would expect based on resonance overlap - the planets spend very little time near the first order mean motion resonances with low $m$ ($P_2/P_1 = 1.5, 1.33, 1.25, 1.2$) nor near the second order resonances with low $m$ ($P_2/P_1 = 1.4, 1.28$). This is because there are large regular islands at these period ratios. The chaotic web has fractionally less volume of phase space near these first and second order resonances, and the orbit spends proportionally less time here. Taken together, these observations strongly suggest that the chaotic web associated with resonance overlap is all connected, as it appears to be in Figure \ref{fig:Illustrate}. Note that as the period ratio grows, the eccentricities of the planets increase as well. At higher eccentricities, though, the chaotic web extends to higher values of the period ratio. This is how an orbit beginning at $P_2/P_1\sim 1.17$ and low eccentricity can reach period ratios as large as 1.7.

Even if a pair of planets will always remain bound to their host star while the star is on the main sequence, we expect the multi-planet systems we observe to be essentially stationary, in that their period ratios should not be varying erratically on short timescales. As a result, we predict that we should not observe systems of two planets with period ratios smaller than the critical period ratio determined by the overlap criterion, even though these systems may be Hill stable.

Our work suggests that the Lagrange instability time of Hill stable close orbits is much shorter than the timescales for ejections. This is in agreement with recent numerical studies of planetary ejection in Hill stable systems \citep{VerasMustill}.  In the region of parameter space studied here, the reason for this may be that the chaotic zone associated with first order mean motion resonances can only extend to $P_2/P_1\sim 2$, and diffusion to period ratios larger than this in a sparse chaotic web will be very slow. It is important to extend the definition of instability to include erratic variations in the semimajor axes especially if using long-term stability as a check for whether or not a given planetary system is stable enough that we would be likely to observe it in its current configuration.

It is interesting that the chaotic zone caused by first order resonance overlap is in fact divided by the Hill boundary for low mass planets (as explained in the following section, and demonstrated in the lower panels of Figure \ref{fig:totalmass}). The Hill boundary corresponds to an invariant surface in the phase space. Orbits on one side cannot cross to the other. The chaotic web is a single chaotic zone in the sense that the mechanism for chaos is the same across it, but it is probably more correct to think of it as two separate chaotic zones, since chaotic orbits cannot explore both sides of the Hill boundary.

\subsection{A Relationship Between the Hill Criterion, Lagrange Instabilities, and Resonance Overlap}
In the previous section, we demonstrated that orbits in the chaotic web caused by first order resonance overlap are unstable, and therefore we can use resonance overlap criteria to determine long-term stability. We now compare the resonance overlap criterion to the Hill criterion and more generally consider the effectiveness of the Hill criterion in the context of the resonance overlap.

In the case of initially circular orbits, two planets are Hill stable if their initial semimajor axes satisfy 
\begin{align}\label{Hill2}
\frac{a_2-a_1}{a_1} \gtrsim 2.4 \epsilon_p^{1/3}.
\end{align}
We have found that for initially circular orbits, first order resonances will overlap and result in a chaotic phase space if $(a_2-a_1)/a_1\lesssim1.46\epsilon_p^{2/7}$. This resonance overlap criterion is absolute in that orbits which fail it, regardless of the eccentricities, are effectively guaranteed to be chaotic (and Lagrange unstable) if they are not protected by the 1:1 resonance.

 As Gladman pointed out, these two boundaries cross, and our work predicts the crossing to occur at a value of $\epsilon_p = (1.46/2.4)^{21}$. Both boundaries are plotted in Figure \ref{fig:IllustrateScaling}. For planets with $\epsilon_p$ greater than $3 \times 10^{-5}$, corresponding to about 10 Earth masses around a solar mass star, the Hill criterion is stricter. However, for smaller planets, the region of complete resonance overlap (where all orbits are expected to be chaotic) extends to larger period ratios $P_2/P_1$ than the Hill boundary.   

Recall, however, that the resonance overlap criterion as calculated here is a minimum criterion for chaos as we have not included the effects of higher order resonances in between the first order resonances. Since the coefficient is taken to the 21st power, even a small increase in the coefficient significantly increases the value of $\epsilon_p$ where the Hill boundary and the overlap boundary are equal.

	Gladman found that for initially circular orbits the Hill criterion was effectively a \textit{necessary} condition for collisional stability of low mass ($\epsilon_p \lesssim 10^{-5}$, arbitrary $\zeta > 1$) planets - though it is only formally sufficient -  but that at higher eccentricities many orbits which failed the criterion were seemingly long-lived. We suggest that the Hill criterion is an effectively necessary condition for stability of low mass planets on initially circular orbits because the region of complete resonance overlap (that which is predicted by the criterion $(a_2-a_1)/a_1\lesssim1.46\epsilon_p^{2/7}$) extends past the Hill boundary and there are no regular regions within it (ignoring the 1:1).  Hence, initially circular orbits which fail the Hill criterion are almost guaranteed to be chaotic, and, as we have seen in the previous section, unstable. For initially circular orbits, the Hill criterion do not depend on $\zeta$ and so this result should be independent of the planetary mass ratio.

\begin{figure}
	\begin{center}
	\includegraphics[width=3.5in]{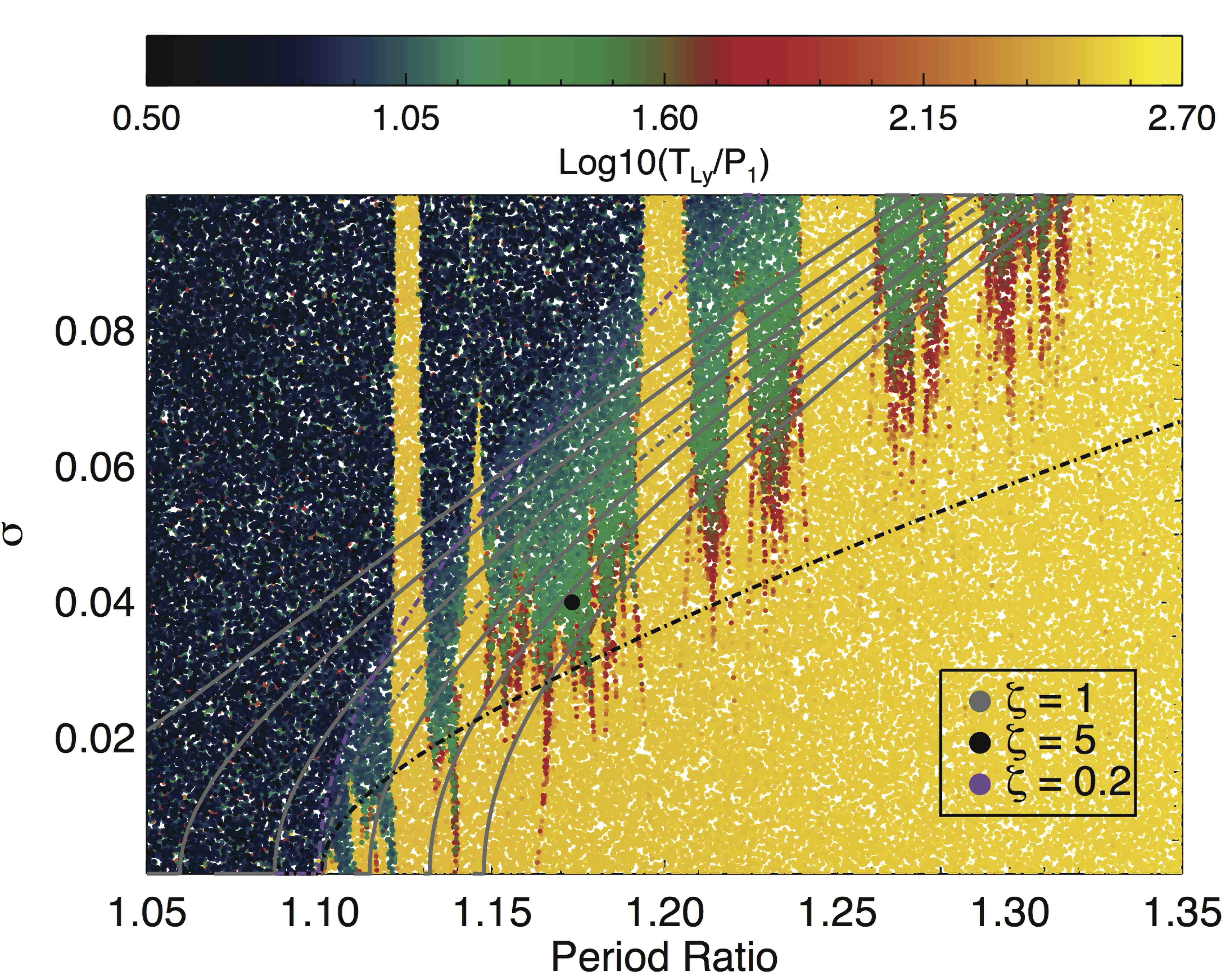}
	\caption{Various contours of $(p/a)/(p/a)\lvert_{crit}$, with $\zeta =1$, overplotted in grey onto the bottom panel of Figure \ref{fig:overlap1}. The dotted line corresponds to the Hill boundary, where $(p/a)/(p/a)\lvert_{crit}=1$. Initial conditions above this line fail the Hill criterion, those below it satisfy the criterion. The contours correspond to values of $(p/a)/(p/a)\lvert_{crit}= 0.9992, 0.9995,0.9998,1.0,1.0002,1.0005,1.0008$ from top to bottom. The large black dot corresponds to the initial condition shown in Figure \ref{fig:evol}. Though the Kepler 36 system does not fit directly onto this plot, that system has a similar total planetary mass relative to the mass of the host star $\epsilon_p$, with $\sigma \lesssim 0.04$  and $P_2/P_1 \sim 1.17$. We also show the critical Hill contour in the case of $\epsilon_p=2\times10^{-5}$ and $\zeta=0.2$ and $\zeta=5$, to show how the Hill contours depend on the mass ratio $\zeta$. }
	\label{fig:dL}
	\end{center}
	\end{figure}

For orbits with arbitrary eccentricities, the criterion for Hill stability can be written as 
	\begin{align}\label{Hill}
	\frac{p}{a} = \frac{-2 (m_\star+m_1+m_2)}{G^2 (m_\star m_1+m_2 m_1+m_\star m_2)^3} L^2 H&> \frac{p}{a} \lvert_{crit},
	\end{align}
	
	where $L$ is the total angular momentum, $H$ is the total orbital energy, $(p/a)\lvert_{crit}$ is a function only of the masses of the bodies, and $p$ and $a$ are the generalized semi-latus rectum and semimajor axis of the problem \citep{Marchal261982,Gladman}. Figure \ref{fig:dL} shows contours of  $(p/a)/(p/a)\lvert_{crit}$ and the underlying chaotic zone as a function of period ratio and $\sigma$ for two equal mass planets with $\epsilon_p = 2\times10^{-5}$.  The dotted contour is the Hill boundary, above this curve all orbits fail the Hill criterion, below it all orbits are Hill stable. 
	
	At larger eccentricities there are still regular regions of phase space which do not satisfy the Hill criterion but will be long-lived. We posit that the Hill criterion is not effectively necessary for stability of (i) moderately eccentric orbits of planets of any $\epsilon_p$ and (ii) higher mass systems ($\epsilon_p \gtrsim \mbox{ few }\times10^{-5}$) on initially circular orbits  since in these cases the region of complete resonance overlap does not encompass the Hill boundary.

	As we have seen in Section \ref{sec:MRcompare}, the mass ratio of the planets does not greatly affect the structure of the chaotic zone and so we show the Hill boundary ($(p/a)/(p/a)\lvert_{crit}=1$) in the cases of $\zeta=0.2$ and $\zeta=5$ in Figure \ref{fig:dL} as well. The Hill criterion depends strongly on the planetary mass ratio $\zeta$.  
%
%
%
%
%
%
%

	We now turn to one of the motivating questions for this work: why is the proximity of an orbit to the Hill boundary (how close $(p/a)/(p/a)\lvert_{crit}$ is to 1) seemingly so important for determining whether or not a given planetary system is Lagrange long-lived? And why is the transition between Lagrange unstable and Lagrange long-lived orbits such a sharp function of this distance from the Hill boundary?	 
			\begin{figure}
	\begin{center}
	\includegraphics[width=3.5in]{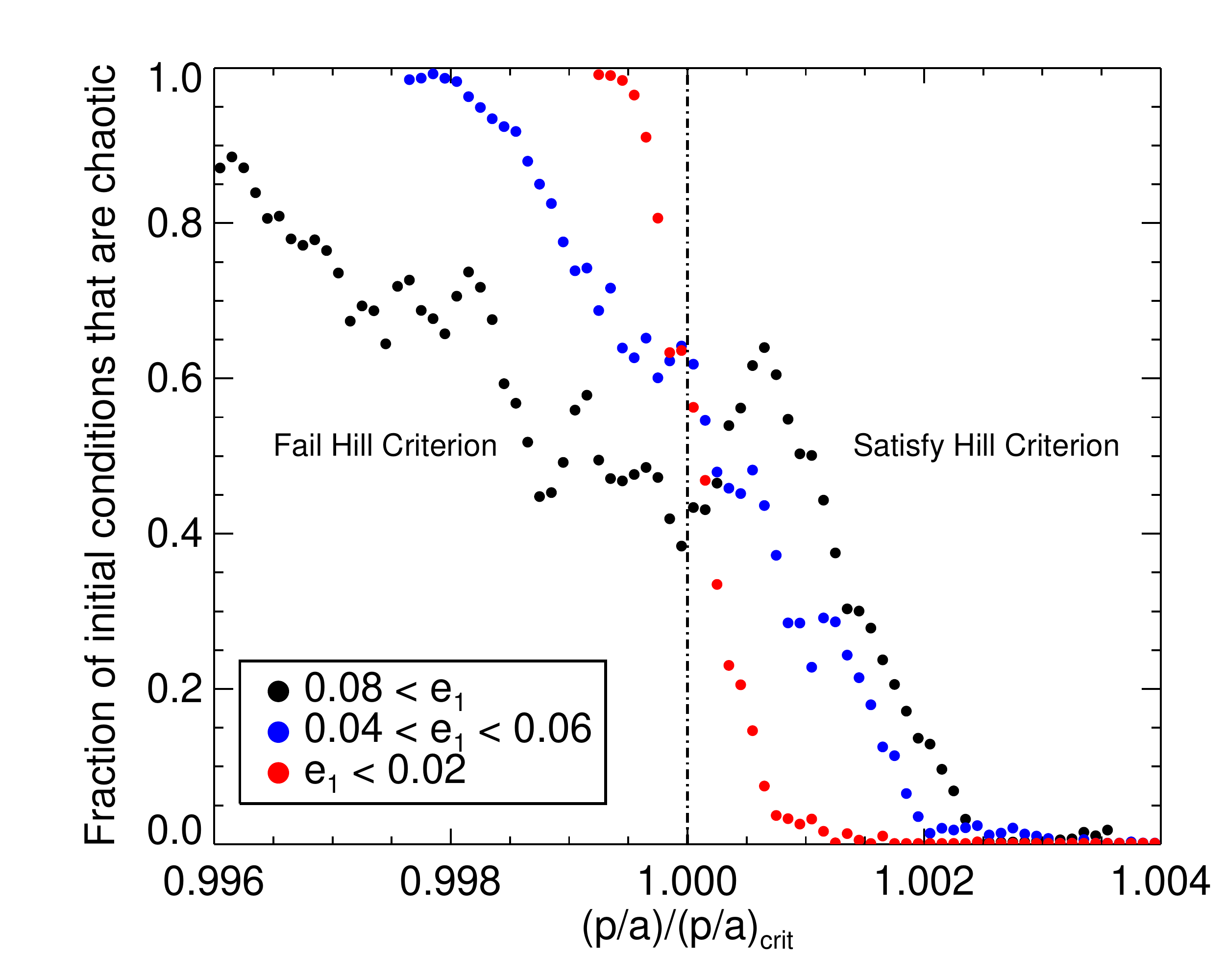}
	\caption{A study of how the chaotic fraction of the phase space changes as a function of $(p/a)/(p/a)\lvert_{crit}$ for equal mass planets. Over 400,000 initial conditions were used in making this plot, taken from the initial conditions used in the three panels of Figure \ref{fig:overlap1} ($\epsilon_p=2\times10^{-5}, \zeta=1 $). A wide range of period ratios and eccentricities were used, as well as different initial orbital angles. Despite this, we see a consistent trend: a sharp transition from an almost entirely chaotic phase space to an entirely regular phase space. The exception is at the highest eccentricities, where regular regions associated with resonances are more important. }
	\label{fig:fraction}
	\end{center}
	\end{figure}
	
	In the case of comparable mass planets, the contours of $(p/a)/(p/a)\lvert_{crit}$ lie approximately parallel to the edge of the chaotic zone, and a small change in the parameter $(p/a)/(p/a)\lvert_{crit}$ moves orbits entirely out of the chaotic region. To explore this further, we determined the fraction of the orbits in each bin of $(p/a)/(p/a)\lvert_{crit}$ which were chaotic using all of the initial conditions in each panel of Figure \ref{fig:overlap1}. These orbits had $\epsilon_p=2\times10^{-5}$ and $\zeta = 1$. The results are shown in Figure \ref{fig:fraction} for three different eccentricity groups. It is clear that at higher eccentricities the fraction of the phase space which is chaotic decreases much less steeply as a function of  $(p/a)/(p/a)\lvert_{crit}$  than it does for lower eccentricities.  
	
	We have shown that chaotic orbits are typically Lagrange unstable, and therefore the sharp transition between an almost entirely chaotic phase space and an almost entirely regular one can naturally explain the observed sharp transition between Lagrange unstable and Lagrange long-lived orbits with $(p/a)/(p/a)\lvert_{crit}$ - in the case of low eccentricities ($\sigma \lesssim 0.05$) and comparable mass planets.  For reference, a stability analysis of the Kepler 36 system ($\epsilon_p \sim 4\times10^{-5}, e_1\sim 0.02, e_2\sim 0.01, \zeta =1.8 $) has shown that the orbits must satisfy  $(p/a)/(p/a)\lvert_{crit} \gtrsim 1.0007$ in order to be long-lived.

	The curves shown in Figure \ref{fig:fraction} will depend on the planetary mass ratio $\zeta$ (because the contours of $(p/a)/(p/a)\lvert_{crit}$ depend strongly on the planetary mass ratio, see Figure \ref{fig:dL}). However, since in the limit of zero eccentricity the Hill boundary is  independent of $\zeta$ (see Equation \eqref{Hill2}), we would predict a similar transition for very low eccentricity orbits regardless of $\zeta$. However, we would not expect a sharp transition to occur in close two planet systems with eccentricities $0.1 \gtrsim \sigma \gtrsim 0.05$, especially for planetary mass ratios far from unity. This is because at larger eccentricities there are regions which fail the Hill criterion yet remain long-lived, as demonstrated by Figure \ref{fig:fraction}, and because contours of $(p/a)/(p/a)\lvert_{crit}$ do not necessarily track the edge of the chaotic zone when $\zeta \not\approx1$.

	We note that in the extreme limit of the elliptic restricted three body problem, $(p/a)\lvert_{crit} \rightarrow 1$ and $(p/a) \rightarrow  1-e^2$, where $e$ is the eccentricity of the massive planet. All orbits fail the Hill criterion (see eqn. (50), \citet{Marchal261982}, or \citet{Nobili}), though this does not imply that all orbits are unstable. In this case, the Hill contours begin to look horizontal on the $(P_2/P_1,\sigma)$ plane as there is no $P_2/P_1$ dependence. As a result, there would be no sharp transition in the Lagrange stability of orbits as a function of $(p/a)/(p/a)\lvert_{crit}$. 
	
	We cannot address the abrupt transition between Lagrange unstable and Lagrange long-lived orbits observed for two planet systems in areas of parameter space different from those studied here, with higher eccentricities ($e \gtrsim 0.1$) and $P_2/P_1\gtrsim2$ \citep{BG1,BarnesandKopparapu,VerasMustill}.  A more complete understanding of a connection between resonance overlap (and Lagrange instability) and the Hill criterion will require more work.

 We hypothesize that instability of close orbits with nearly circular orbits ($\sigma \lesssim 0.1$) is caused by first order resonance overlap, either between first order resonances directly, or indirectly between first and second order resonances. However, the first order resonances are crucial to the overlap because they effectively connect regions of phase space with $P_2/P_1 \sim 1$ to regions of phase space with $P_2/P_1 \sim 1.5-2$; the first order resonance provide the backbone structure of the chaotic web.  An analytic derivation of the boundary of the chaotic zone resulting from first and second order resonance overlap, as a function of eccentricity and period ratio, is beyond the scope of this paper. However, we predict that a derivation for the extent of this chaotic zone would serve as an effective Lagrange criterion for stability of systems of two planets. This is left to future work.

\section{CONCLUSION}
The Hamiltonian governing the dynamics of two planets near first order mean motion resonance can be reduced to a one degree of freedom system when the orbital eccentricities and inclinations are small \citep{Sessin,WisdomCtB,Lemaitre}. We have used the Hamiltonian in this reduced form to study the structure of the first order resonances and to derive their widths as a function of orbital elements of the planets. These widths are predicted to be independent of the planetary mass ratio $\zeta = m_1/m_2$. Using these analytically determined widths, we have derived a resonance overlap criterion for two massive planets in the cases of initially circular orbits and initially eccentric orbits. For the former case, we find that the first order mean motion resonances will overlap when $(a_2-a_1)/a_1<1.46\epsilon_p^{2/7}$, where $\epsilon_p$ is the total mass of the planets, relative to the mass of the host star. This implies that approximately all orbits failing this criterion (even those that are eccentric) will be chaotic unless they are protected by the 1:1 resonance.

These analytic results were compared extensively to numerical integrations, which confirmed the general predictions of the analytic theory as to the locations and widths of the first order resonances. These integrations demonstrated that the derived overlap criterion for initially circular orbits closely tracks the edge of the numerically determined chaotic zone at zero eccentricity. However, it is clear that chaotic structure of phase space at higher eccentricities and larger period ratios is partially due to higher order mean motion resonances between the first order resonances. As a result, the analytic estimate of the width of the chaotic zone caused by first order resonance overlap as a function of eccentricities may not be very applicable in practice. 

Our numerical investigations show that the chaotic structure caused by first (and higher) order resonance overlap forms a ``chaotic web" in phase space, with regular regions appearing only deep in low order resonances. We have found that the structure of this chaotic web is indeed independent of the mass ratio of the planets $\zeta$. Though this is what we expected based on the first order resonance theory, we find it surprising as it suggests that the widths of second order resonances are approximately independent of $\zeta$ in the regime $P_2/P_1 \lesssim 2, \sigma \lesssim 0.1$. Numerical integrations show that (i) chaotic orbits within the chaotic web are generally Lagrange unstable on short timescales, (ii) the timescale to exhibit instability grows the closer the orbit is to the boundary of the chaotic web, and (iii) the timescale to exhibit instability is apparently not related to the Lyapunov time of the orbits. Finally, we have demonstrated that chaotic orbits explore the entire extent of the chaotic web (and therefore the chaotic web is likely a single connected chaotic zone).

These studies demonstrate that resonance overlap criteria can be used as stability criteria. We have shown that for systems with $\epsilon_p \lesssim 10 M_\oplus$ and initially circular orbits the resonance overlap criterion is a more restrictive stability criterion than the Hill criterion. This led us to an explanation for  Gladman's numerical result that the Hill criterion is effectively a necessary criterion for stability of low mass initially circular two planet systems. More generally, our work suggests that orbital instabilities of close planets are generated by overlap of first (and higher) order mean motion resonances. Therefore we believe that a more extensive resonance overlap criterion which incorporated first and second order resonances could be used as a Lagrange stability criterion for observed pairs of planets (and possibly be a better indicator of stability than the Hill criterion).

 Lastly, we have taken some initial steps towards understanding the observed relationship between the Hill criterion and measured Lagrange instability boundaries.  In the special case of approximately equal mass planets,  contours of the distance from the Hill boundary appear to lie parallel to the edge of the chaotic zone caused by resonance overlap. This naturally produces a sharp transition from an entirely chaotic (and unstable) phase space to an almost entirely long-lived (and regular) phase space for low eccentricity orbits ($e\lesssim 0.05$). However, it is clear there is still much to be done to understand the connection between Lagrange instability and the Hill criterion for more eccentric orbits and for systems of two planets with very unequal masses.

\acknowledgments
K.M.D. acknowledges support from an NSF Graduate Research Fellowship. M.J.P. acknowledges supports from the NASA Kepler Participating Scientists Program and from the NASA Origins of Solar Systems Program. We would also like to thank the anonymous referee for reading the text so closely and for many helpful suggestions. 

\bibliographystyle{apj}
\bibliography{ms}
  \renewcommand{\theequation}{A.\arabic{equation}}    
  \setcounter{equation}{0}  
    \renewcommand{\thesubsection}{A-\arabic{subsection}}   
     \setcounter{subsection}{0}  
        \renewcommand{\thefigure}{A.\arabic{figure}}   
          \setcounter{figure}{0}  
  
  \section*{APPENDIX}  

\subsection{Transformation of Coordinates}
The numerical integrations evolve the equations of motion for the full, exact, Hamiltonian, while the theory predicts the extent of the resonances in terms of a different set of canonical variables. In order to  compare the two sets of results, one must perform the sequence of canonical transformations between the two set of variables. This requires removing the short period terms via a generating function so that they do not appear in the averaged Hamiltonian. It is equivalent to averaging the full Hamiltonian over the fast angle $\lambda_i$, at least at first order in $\epsilon$, because the resulting Hamiltonian has the same functional form. 

The deviation between the averaged and full trajectories is of order $\epsilon$, and hence the canonical transformation between the two sets of variables should be very near the identity transformation. In most cases, it is entirely negligible. However, that is not always true when the two orbits are close to each other.

The full Hamiltonian near the $m$:$m+1$ resonance can be written to first order in eccentricities as
\begin{align}\label{sigh}
 H &= H_0(\mathbf{\Lambda})+\epsilon_1 H_1(\mathbf{\Lambda},\mathbf{x},\mathbf{\lambda},\mathbf{y}) \nonumber \\
 H_0 & = -\frac{\mu_1}{2\Lambda_1^2} -\frac{\mu_2}{2\Lambda_2^2}  \nonumber \\
 \epsilon_1 & = \frac{m_1}{m_\star} \nonumber \\
 H_1 & = -\frac{\mu_2}{\Lambda_2^2}\bigg[ \frac{f_{m+1,27}(\alpha_{res})}{\sqrt{\Lambda_1}} (\cos{\theta_m} x_1 -\sin{\theta_m} y_1) +\frac{f_{m+1,31}(\alpha_{res})}{\sqrt{\Lambda_2}} (\cos{\theta_m} x_2 -\sin{\theta_m} y_2) + \nonumber \\
 &\sum_{k>0,k \neq m} \frac{f_{k+1,27}(\alpha_{res})}{\sqrt{\Lambda_1}} (\cos{\theta_k} x_1 -\sin{\theta_k} y_1) +\frac{f_{k+1,31}(\alpha_{res})}{\sqrt{\Lambda_2}} (\cos{\theta_k} x_2 -\sin{\theta_k} y_2) +   \sum_{j>0}f_{j,1}(\alpha_{res})\cos{\phi_j} \bigg] 
\end{align}
where all Laplace coefficients are evaluated at the resonance location $\alpha_{res} = (m/(m+1))^{2/3}$. The variables $x$ and $y$ are the same canonical eccentricity polar coordinates introduced in Equation \eqref{polar}. For notational simplicity, we have also defined $\theta_i = (i+1) \lambda_2 - i\lambda_1$ and $\phi_j = j(\lambda_2-\lambda_1)$.

Again note that $f_{2,31}$ must include an indirect contribution of $-2\alpha_{res}$, and that $f_{1,1}$ includes an indirect contribution of $-\alpha_{res}$. We have ignored two terms in the disturbing function which come from the indirect part. These terms are proportional to $e_1 \alpha_{res}$ and are independent of the Laplace coefficients. They depend on the combinations of the fast angles $\lambda_2-2\lambda_1$ and $\lambda_2$. We do not prove it now, but these terms will be negligible to the transformation and so we ignore them. We'll return to this point later. 

This expression can be written more simply as 
\begin{align}
H &= H_{Kepler}(\Lambda_1,\Lambda_2)+  \epsilon_1H_{R,FO,m }(\mathbf{z})+ \sum_{k>0,k \neq m}  \epsilon_1 H_{SP,FO,k}(\mathbf{z})+ \sum_{k>0}   \epsilon_1H_{SP,circ,k}(\mathbf{z})
\end{align}
where $R$ implies resonant, $SP$ implies short period terms, $FO$ implies a first order term,  $circ$ implies the term appears at zero eccentricity, and $\mathbf{z}$ denotes the phase space variables.

We seek a generating function of Type 2 which determines a near-identity transformation that removes the short period terms. We will write this generating function as
\begin{align}
F_2(\lambda_i,y_i,\tilde{\Lambda}_i,\tilde{x}_i) &= \lambda_i \tilde{\Lambda}_i + y_i\tilde{x}_i + \epsilon_1 \it{f}(\lambda_i,y_i,\tilde{\Lambda}_i,\tilde{x}_i) 
\end{align}
where $\it{f}$ is yet to be determined.

Here the variables with tildes are the averaged canonical coordinates and momenta, while those without are the exact physical phase space coordinates. The old and new coordinates are related as
\begin{align}\label{transform}
\Lambda_i & = \frac{\partial F_2}{\partial \lambda_i} = \tilde{\Lambda}_i+\epsilon_1\frac{\partial \it{f}}{\partial \lambda_i} \nonumber \\
x_i & = \frac{\partial F_2}{\partial y_i} = \tilde{x}_i+\epsilon_1\frac{\partial \it{f}}{\partial y_i} \nonumber \\
\tilde{\lambda}_i & = \frac{\partial F_2}{\partial \tilde{\Lambda}_i} = \lambda_i+\epsilon_1\frac{\partial \it{f}}{\partial \tilde{\Lambda}_i} \nonumber \\
\tilde{y}_i & = \frac{\partial F_2}{\partial \tilde{x}_i} = y_i+\epsilon_1\frac{\partial \it{f}}{\partial \tilde{x}_i} 
\end{align}

The new (averaged) Hamiltonian is $\tilde{H} (\tilde{\mathbf{z}}) = H(\mathbf{z}(\tilde{\mathbf{z}}))$ since the generating function is time independent. To first order in $\epsilon_1$,
\begin{align}
\tilde{H}(\tilde{\mathbf{z}}) &= H(\mathbf{z}(\tilde{\mathbf{z}})) \nonumber \\
&= H_0(\tilde{\Lambda}_1+\epsilon_1\frac{\partial \it{f}}{\partial \lambda_1},\tilde{\Lambda}_2+\epsilon_1\frac{\partial \it{f}}{\partial \lambda_2}) +    \epsilon_1 H_{R,FO,m }(\tilde{\mathbf{z}})+\sum_{k>0,k \neq m}   \epsilon_1 H_{SP,FO,k}(\tilde{\mathbf{z}})+ \sum_{k>0}  \epsilon_1 H_{SP,circ,k}(\tilde{\mathbf{z}}) \nonumber \\
& = H_0(\tilde{\Lambda}_1, \tilde{\Lambda}_2) +  \epsilon_1 \bigg( \frac{\partial H_0(A,B)}{\partial A} \bigg \lvert_{(A,B) = (\tilde{\Lambda}_1, \tilde{\Lambda}_2)} \frac{\partial \it{f}}{\partial \lambda_1} (\tilde{\mathbf{z}}) +  \frac{\partial H_0(A,B)}{\partial B} \bigg\lvert_{(A,B) = (\tilde{\Lambda}_1, \tilde{\Lambda}_2)} \frac{\partial \it{f}}{\partial \lambda_2} (\tilde{\mathbf{z}}) \bigg)+ \nonumber \\
& \epsilon_1H_{R,FO,m }(\tilde{\mathbf{z}})+\sum_{k>0,k \neq m}   \epsilon_1H_{SP,FO,k}(\tilde{\mathbf{z}})+ \sum_{k>0} \epsilon_1 H_{SP,circ,k}(\tilde{\mathbf{z}}) \nonumber \\
 & =  H_0(\tilde{\Lambda}_1, \tilde{\Lambda}_2) + \epsilon_1 (n_{0,1} \frac{\partial \it{f}}{\partial \lambda_1} +n_{0,2} \frac{\partial \it{f}}{\partial \lambda_2}) +\epsilon_1 H_{R,FO,m }(\tilde{\mathbf{z}})+\nonumber \\
 &\sum_{k>0,k \neq m}  \epsilon_1 H_{SP,FO,k}(\tilde{\mathbf{z}})+\sum_{k>0}  \epsilon_1H_{SP,circ,k}(\tilde{\mathbf{z}})
\end{align}
where $n_{0,i}$ is the unperturbed mean motion of planet $i$. Note that $\mathbf{z}$ and $\tilde{\mathbf{z}}$ are interchangeable in all terms of order $\epsilon_1$.

Now, we would like this averaged Hamiltonian to be equal to
\begin{align}
\tilde{H} & =   H_0(\tilde{\Lambda}_1, \tilde{\Lambda}_2) + \epsilon_1 H_{R,FO,m }(\tilde{\mathbf{z}})
\end{align}
i.e. with all short period terms removed. We must choose the function $f$ to satisfy the relation:
\begin{align}\label{satisfy}
 (n_{0,1} \frac{\partial \it{f}}{\partial \lambda_1} &+n_{0,2} \frac{\partial \it{f}}{\partial \lambda_2})  = - \sum_{k>0,k \neq m}  H_{SP,FO,k}- \sum_{k>0} H_{SP,circ,k}
\end{align}

The solution is given as
\begin{align}
f &=  -\frac{\mu_2}{\tilde{\Lambda}_2^2}\bigg[ \sum_{k>0,k \neq m} A_{0,k} \bigg(\frac{f_{k+1,27}(\alpha_{res})}{\sqrt{\tilde{\Lambda}_1}} (\sin{\theta_k} \tilde{x}_1 +\cos{\theta_k} y_1) +\frac{f_{k+1,31}(\alpha_{res})}{\sqrt{\tilde{\Lambda}_2}} (\sin{\theta_k} \tilde{x}_2 +\cos{\theta_k} y_2)\bigg) + \nonumber \\
&  \sum_{j>0}B_{0,j} f_{j,1}(\alpha_{res})\sin{\phi_j}  \bigg]
\end{align}

One can write this down almost immediately because the short period Hamiltonian can be written entirely in terms of cosine functions, i.e. in the form of $H_{SP} = \sum_i C_i \cos{(\beta_i)}$.  This is true of the combinations $x_i \cos{\theta}-y_i \sin{\theta}$ which can be written as $\sqrt{2 P_i} \cos{(\theta+p_i)}$. The function $f$ is simply the same function, scaled by constant coefficients, but with the cosines switched to sines: $f = \sum_i C'_i \sin{(\beta_i)}$. 
Then, 
\begin{align}\label{derivofF}
\frac{\partial \it{f}}{\partial \lambda_1} & =   \sum_{k>0,k \neq m}  A_{0,k} \frac{\partial \theta_k }{\partial \lambda_1} H_{SP,FO,k} +  \sum_{k>0} B_{0,k}H_{SP,circ,k}  \frac{\partial \phi_k }{\partial \lambda_1} \nonumber \\
 & =  \sum_{k>0,k \neq m} A_{0,k} (-k) H_{SP,FO,k} +  \sum_{k>0} B_{0,k} H_{SP,circ,k} (-k) \nonumber \\
\frac{\partial \it{f}}{\partial \lambda_2} & =  \sum_{k>0,k \neq m}  A_{0,k} \frac{\partial \theta_k }{\partial \lambda_2} H_{SP,FO,k} +  \sum_{k>0} B_{0,k}H_{SP,circ,k}  \frac{\partial \phi_k }{\partial \lambda_2} \nonumber \\
 & =  \sum_{k>0,k \neq m} A_{0,k} (k+1) H_{SP,FO,k} +  \sum_{k>0} B_{0,k} H_{SP,circ,k} (k) 
\end{align}
Using the derivatives of $f$ given in Equation \eqref{derivofF} in Equation \eqref{satisfy} then results in an equation for the $A_{0,k}$ and $B_{0,k}$:
\begin{align}\label{a0b0}
 \sum_{k>0,k \neq m} &(A_{0,k} (-k n_{0,1}+ (k+1) n_{0,2}) +1) H_{SP,FO,k}  + \sum_{k>0} (B_{0,k} ( -k n_{0,1} +k n_{0,2}) +1) H_{SP,circ,k}  =0 \nonumber \\
A_{0,k} & = \frac{-1}{(k+1) n_{0,2}-k n_{0,1}}  = \frac{1}{(k+1)n_{0,2}(X_k-1)} \nonumber \\
B_{0,k} & = \frac{-1}{k(n_{0,2}-n_{0,1})}  = \frac{1}{k n_{0,2}(P_2/P_1-1)} \nonumber \\
X_k & = \frac{k P_2}{(k+1) P_1}
\end{align}
where the first equation must hold for each $k$ individually.

To find the transformation between the variables, we use Equation \eqref{transform}. First, let us determine how to map initial period ratios used in the numerical integrations to ``averaged" period ratios.
The $\it{real}$ period ratio of the planets is given by $P_2/P_1 \bigg \lvert_{R}= (\Lambda_2/\Lambda_1)^{3}*(m_1/m_2)^{3}$, while the averaged period ratio is given by $P_2/P_1 \bigg \lvert_{A}= (\tilde{\Lambda}_2/\tilde{\Lambda}_1)^{3}*(m_1/m_2)^{3}$. 
Therefore,
\begin{align}
P_2/P_1 \bigg \lvert_{R} &\approx P_2/P_1 \bigg \lvert_{A}\bigg[1+3\epsilon_1\frac{\partial \it{f}}{\partial \lambda_2} /\Lambda_2 \bigg] \bigg[1+3 \epsilon_1\frac{\partial \it{f}}{\partial \lambda_1}/\Lambda_1 \bigg] ^{-1} \nonumber \\
& \approx P_2/P_1 \bigg \lvert_{A}\bigg[1-3 \epsilon_1\frac{\partial \it{f}}{\partial \lambda_1} /\Lambda_1+  3 \epsilon_1 \frac{\partial \it{f}}{\partial \lambda_2} /\Lambda_2\bigg]
\end{align}
The required derivatives of $f$ are:
\begin{align}
\frac{\partial \it{f}}{\partial \lambda_1} /\Lambda_1 & =  \sum_{k>0,k \neq m} \frac{k }{k+1} \frac{\Lambda_2}{\Lambda_1} \frac{1}{X_k-1} \bigg[f_{k,27} e_1 \cos{(\theta_k-\varpi_1)}+f_{k,31} e_2 \cos{(\theta_k-\varpi_2)}\bigg]+   \sum_{k>0} \frac{f_{k,0}}{P_2/P_1-1} \frac{\Lambda_2}{\Lambda_1}\cos{\phi_k} \nonumber \\
\frac{\partial \it{f}}{\partial \lambda_2} /\Lambda_2 & =  \sum_{k>0,k \neq m} - \frac{1}{X_k-1} \bigg[ f_{k,27} e_1 \cos{(\theta_k-\varpi_1)}+ f_{k,31} e_2 \cos{(\theta_k-\varpi_2)}\bigg] +   \sum_{k>0} -\frac{f_{k,0}}{P_2/P_1-1} \cos{\phi_k}
\end{align}

so that
\begin{align}
P_2/P_1 \bigg \lvert_{R} &= P_2/P_1 \bigg \lvert_{A}\times \nonumber \\
&\bigg\{1-3\epsilon_1  \sum_{k>0,k \neq m} \bigg[\frac{k }{k+1} \frac{\Lambda_2}{\Lambda_1}+1\bigg]\frac{1}{X_k-1} \bigg[ f_{k,27} e_1 \cos{(\theta_k-\varpi_1)}+ f_{k,31} e_2 \cos{(\theta_k-\varpi_2)}\bigg]  \nonumber \\
&- 3\epsilon_1 \sum_{k>0} \frac{f_{k,0}}{P_2/P_1-1} \cos{\phi_k}(1+\Lambda_2/\Lambda_1) \bigg\}
\end{align}
where $\Lambda_2/\Lambda_1 = m_2/m_1(P_2/P_1)^{1/3}$. Again, we can neglect the difference between real and averaged coordinates at order $\epsilon_1$.

There are two reasons why this shift is not negligible. The first is that the period ratio of the planets we are considering is near unity, so the divisors in the coefficients $A_{0,k}$ and $B_{0,k}$ are small. Though the functions proportional to $A_{0,k}$ also are linear in the eccentricities, the functions $f_{k+1,27}$ and $f_{k+1,31}$ are larger when evaluated at $\alpha$ near unity. 

In order to remove a short period term in the Hamiltonian which is periodic in the angle $p \lambda_2-q\lambda_1$, the transformation must include a term proportional to the coefficient of that term and divided by $p n_{0,2}-q n_{0,1}$. This is why the two indirect terms period in the combinations  $\lambda_2-2\lambda_1$ and $\lambda_2$ are negligible. They appear at first order in eccentricities, but with divisors of $n_{0,2}-2 n_{0,1}$ and $n_{0,2}$, respectively, neither of which are small. Moreover, they do not depend on the Laplace coefficients, which makes them even smaller compared to the terms in the transformation coming from the direct terms of first order in eccentricities (and low $k$).

We only require the transformation from real coordinates to average coordinates at the initial time of the numerical integrations. Hence, for all parameters we substitute their initial values to determine the mapping between real and averaged variables. For example, a numerical simulation using $e_2=0$ and all of the angles equal to zero except for $\lambda_1 = M_1$ would require a transformation of
\begin{align}\label{shiftFormula}
P_2/P_1 \bigg \lvert_{R} &= P_2/P_1 \bigg \lvert_{A}\times \nonumber \\
&\bigg\{1-3\epsilon_1  \sum_{k>0,k \neq m} \bigg[\frac{k }{k+1} \frac{\Lambda_2}{\Lambda_1}+1\bigg]\frac{1}{X_k-1} \bigg[  f_{k,27} e_1(0) \cos{k M_1}\bigg] -  3 \epsilon_1 \sum_{k>0} \frac{f_{k,0}}{P_2/P_1-1} \cos{k M_1}(1+\Lambda_2/\Lambda_1) \bigg\}
\end{align}

As the period ratio of the planets grows closer to unity, the convergence rates of the two sums are much slower. When $e_1$ is small, the most important contribution is from the second sum. Each term will have the same sign as $\cos{kM_1}$ does since $f_{k,0}>0$ and $P_2/P_1>1$. It is only when $M_1=0$ that these terms all have the same sign and add coherently. In this case, the period ratio in ``real" variables is smaller than the period ratio in averaged variables by a maximum amount. When $M_1$ is nonzero, the shift is much smaller since terms will not add coherently. This behavior is reflected in Figures \ref{fig:overlap1}, \ref{fig:Illustrate}, and \ref{fig:totalmass}. This incoherent adding of terms occurs even for the sum that appears linearly in $e_1(0)$, and so the shift is should be small regardless of $e_1(0)$ if $M_1 \neq0$. 

The other transformations are straightforward to derive given the generating function. Of interest to us in particular is the transformation of the eccentricities, since we plot the widths of the resonances as a function of eccentricities. We will just give the result here for the case where $e_2=0$ and all angles equal to zero. In this special case, 
\begin{align}
\frac{\partial f}{\partial \tilde{x}_i} &\propto \sin{((k+1)\lambda_2-k\lambda_1)} = 0 \nonumber \\
\frac{\partial f}{\partial y_i} &\propto \cos{((k+1)\lambda_2-k\lambda_1)} = 1
\end{align}
So to first order in $\epsilon_1$, $y_i = \tilde{y}_i$, and
\begin{align}\label{Xcorrect}
x_1 &= \tilde{x}_1 +\epsilon_1  \frac{\Lambda_2}{\sqrt{\Lambda_1}} \sum_{k>0,k \neq m} \frac{\lvert f_{k,27} \rvert}{(X_k-1)(k+1)} \nonumber \\
x_2 &= \tilde{x}_2 -\epsilon_1 \sqrt{\Lambda_2} \sum_{k>0,k \neq m} \frac{f_{k,31} }{(X_k-1)(k+1)} 
\end{align}
Note that these expressions are already of order $(e)$, so no $e$ appears in the sum.

This correction is negligible, however. Although the terms in the sums in Equation \eqref{Xcorrect} look very similar to those in the correction to the period ratio, they are weighted by a factor of $1/(k+1)$, and this makes the correction much smaller. Taken together with the relationship $y_i = \tilde{y}_i$, this implies that both eccentricities and longitudes of periastron are not affected greatly by the transformation. Moreover, this should be true regardless of if $e_2$ is nonzero or if the angles are nonzero because the weighting factor $1/(k+1)$ will always appear.

Finally, we check the correction to the mean longitudes $\lambda_i$. We do this only for our nominal case, with $e_2=0$ and all other angles equal to zero. In this case, $y_i = 0$ and all the $\phi_j$ and $\theta_k$ are equal to zero, so all derivatives with respect to $\tilde{\Lambda}_i$ are zero. In other words, $\lambda_i = \tilde{\lambda}_i$ in this case. 
%
%
%
%
%
%
%
%
%

\subsection{Effect of Integration Time on the Estimated Lyapunov Times}
\begin{figure}
	\begin{center}
	\includegraphics[width=3.5in]{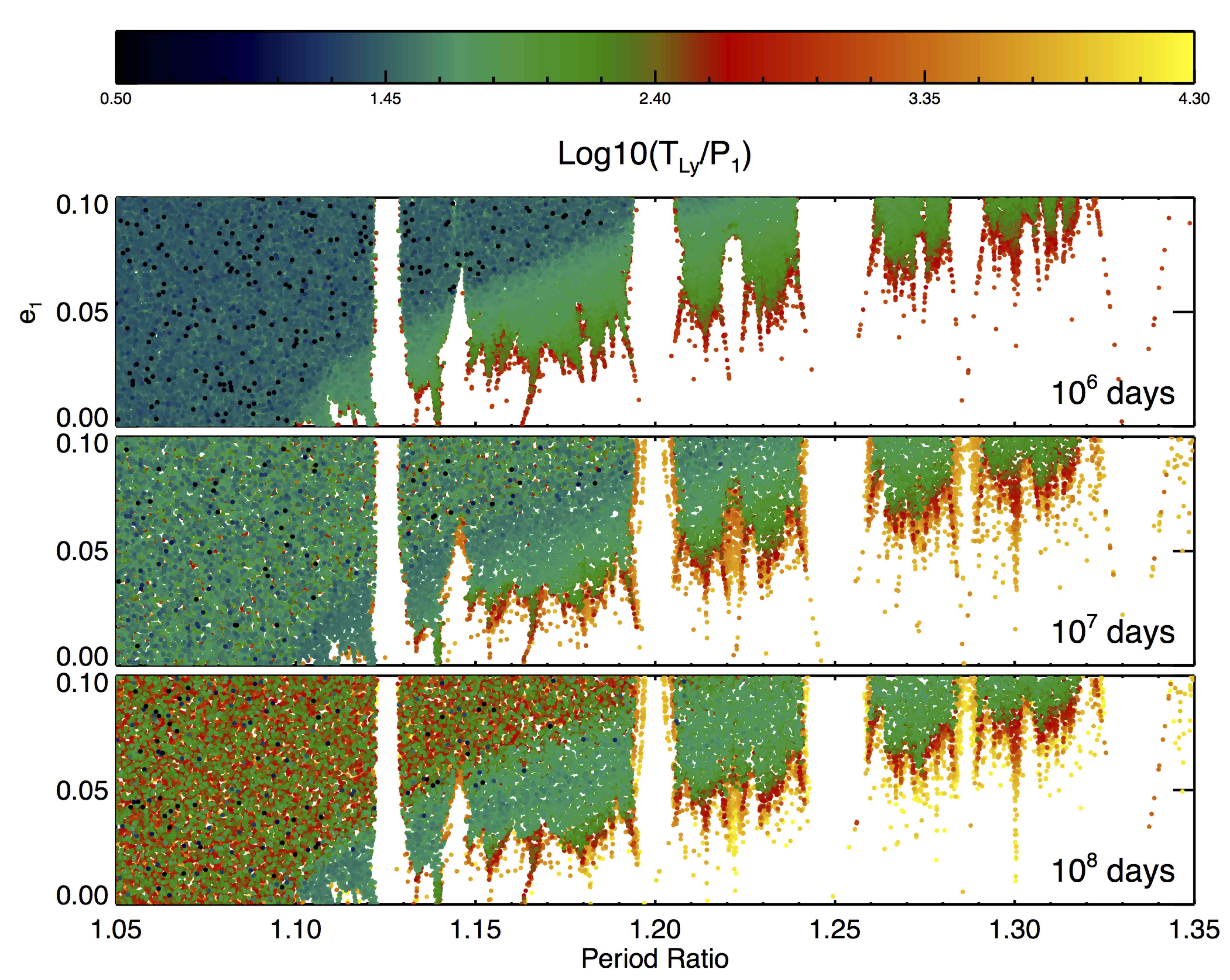}
	\caption{Effect of the integration time on the estimated Lyapunov times. \textit{Top:} $10^6$ day integration, with $e_2$ initially zero, all angles set to zero except $M_1 = \pi/2$, $\zeta=1$, and $\epsilon_p = 2\times10^{-5}$. Only the chaotic orbits are shown in color. White regions contain quasiperiodic orbits. \textit{Middle:} the same initial conditions, but evolved only for $10^7$ days. \textit{Bottom:} the same initial conditions, but evolved for $10^8$ days. $10^6$ days was our standard integration length.  }
	\label{fig:Convergence}
	\end{center}
	\end{figure}

Although numerical integrations can determine if an orbit is chaotic, they cannot prove that an orbit is quasiperiodic. Longer integrations provide better estimates of the Lyapunov time and give a better idea of what orbits could be quasiperiodic. We performed a set of numerical integrations only varying the total integration time order to check the convergence of our estimate of the Lyapunov time. In Figure \ref{fig:Convergence} we show the structure of the chaotic zone, in terms of estimated Lyapunov times, for integration lengths of $10^6$, $10^7$, and $10^8$ days. The three panels agree very well, and give us confidence that $10^6$ days is an adequate amount of time to reliably estimate Lyapunov times for these types of orbits. 

There are two minor differences between the three sets. First, we see that some of the orbits around the second order resonances are chaotic, though the shorter integration did not flag them as such. This is not uncommon; the tangent vector corresponding to a chaotic orbit exponentially grows at different rates, depending on the region of the chaotic zone. For example, the local estimate of the Lyapunov time can be very long when the orbit is near invariant curves associated with resonances. 

Secondly, the estimate of the Lyapunov time slightly shifts to longer times during longer integrations. Our hypothesis is that this is caused by instabilities which set in on timescales of  $10^7$ or $10^8$ days (but not  $10^6$ days). For example, orbits which undergo strong scattering events can lead to a system of a single bound planet and a second planet on a hyperbolic orbit. Though the orbits would still be formally chaotic, the motion is closer to regular after the scattering (the perturbations are weaker). We tracked the log of the norm of the tangent vector for a system which underwent scattering and indeed found that the tangent vector stopped growing exponentially after the scattering event, and hence longer integrations of this orbit will lead to longer and longer estimates of the Lyapunov time.  

Even in cases without planetary ejection, Lagrange instabilities cause the planets to become more and more widely spaced. Again, the perturbations between the planets decreases. We also tracked the log of the norm of the tangent vector for a system undergoing Lagrange instabilities, and found that the estimated Lyapunov time grew as the period ratio of the pair grew further from unity. The longer integrations indicate a longer Lyapunov time because a larger fraction of the integration time is spent at larger period ratios.

\subsection{Results for a Case with $e_2$ Nonzero}
 All of our tests used $e_2=0$,  but the analytic theory predicts that only the weighted eccentricity, and not $e_1$ and $e_2$ individually, matters for the structure of the first order resonances at low eccentricity. We tested this numerically for a single case with $e_2=0.02$. These integrations had all angles initially set to zero except $\varpi_1=\pi/2$. In this case, $\Delta \varpi =\pi/2$ and so $\sigma = \sqrt{e_2^2+e_2^2/R^2}$. 
  
\begin{figure}
	\begin{center}
	\includegraphics[width=3.5in]{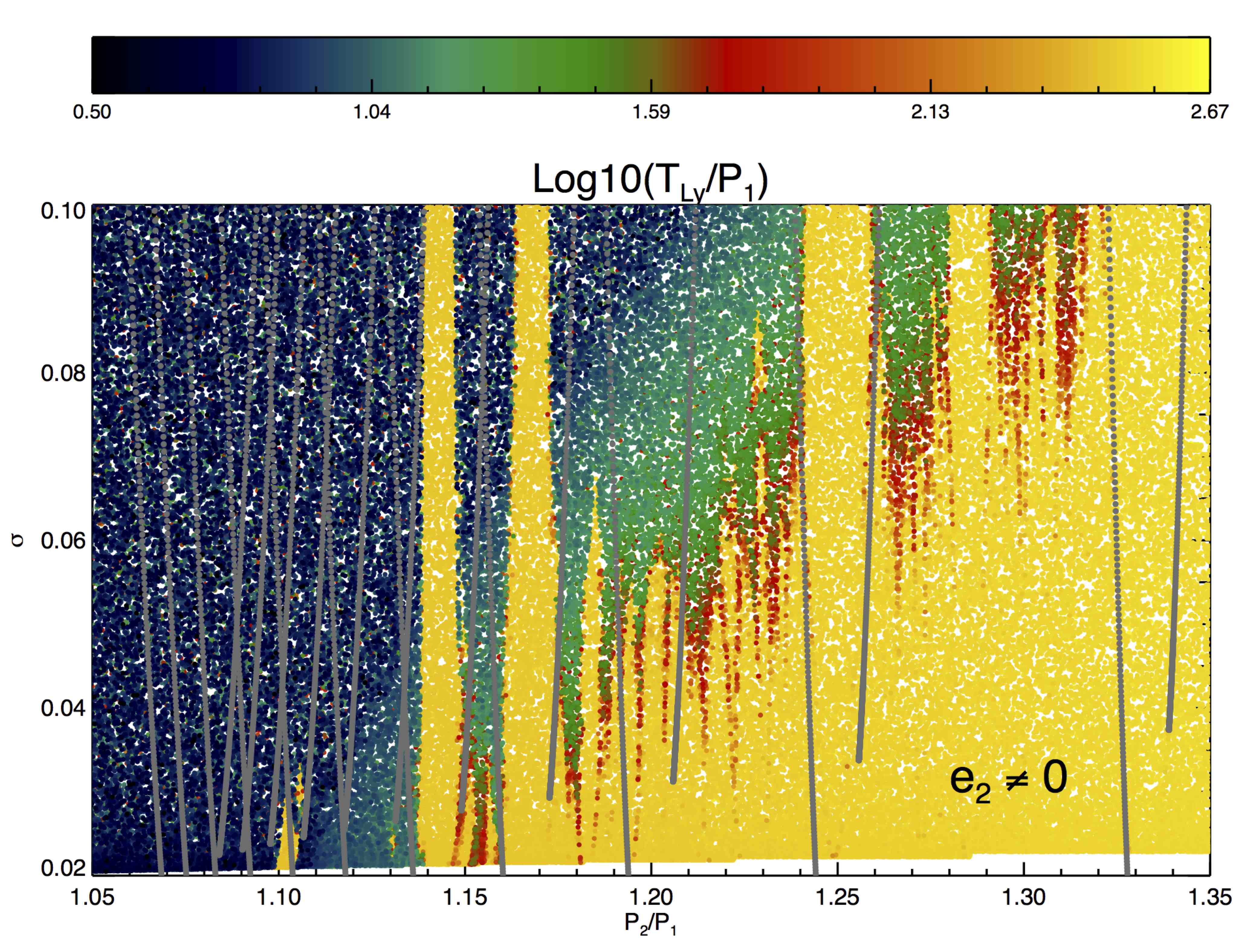}
	\caption{The chaotic structure of phase space when $e_2$ is nonzero and fixed at 0.02. It does not matter what $e_1$ and $e_2$ are individually, only the weighted eccentricity $\sigma$ matters in determining the widths of the first order resonances. This figure shows the chaotic regions of phase space for a set of integrations where $e_1$ varied from 0.0 to 0.1. Besides $e_1$, only the period ratio of the planets varied. The masses of the planets were set as $\epsilon_1=\epsilon_2 = 10^{-5}$. These result shown are what we expect if $\sigma$ is indeed the relevant parameter.  }
	\label{fig:e2}
	\end{center}
	\end{figure}

 For these choices of angles, one can show that the generalized longitude of pericenter $\psi = \arctan{\{s_1/r_1\}} = \arctan{\{R e_1/e_2\}}$. Given the values of $e_1$ and $e_2$, $\psi$ ranges from 0 to $\sim 79^\circ$. The value of the resonant angle $\phi$, then, is $\phi = -m\varpi_1+\psi = -m\pi/2+\psi$. Recall that $\phi=\pi$ corresponds to the center of the resonance, and so if the resonance overlap criterion is not satisfied resonances with $\phi=\pi$ may appear as regular regions (whether or not they do also depends on $e_1$ and $\epsilon_p$). When $\phi$ is closer to zero the orbits are more likely to be chaotic (especially for period ratios closer to unity). Given the range of $\psi$, it is clear that only resonances with $m= 2,6,\dots$, or $m=3,7,\dots$ should appear regular. These resonances are the 3:2, the 4:3, the 7:6, and the 8:7. The 3:2 does not lie within the range of period ratios shown in Figure \ref{fig:e2}, but the others do, they are the only regular first order resonant regions that appear. The predicted seperatrices match up well with the boundaries of these regular regions.

\clearpage

\end{document}